\documentclass[letterpaper,journal,final,twocolumn]{IEEEtran}

\usepackage{amssymb}
\usepackage{amsmath}
\usepackage{amsfonts}                                
\usepackage[latin1] {inputenc}
\usepackage{enumerate}
\usepackage{mathrsfs}
\usepackage{tabulary}
\usepackage{cite}
\usepackage{psfrag}
\usepackage{cite}
\usepackage{graphicx}          
\usepackage{color}      
\usepackage{epsfig} 
\usepackage{epstopdf}
\usepackage{times} 
\def\qedp{\hspace*{\fill}~{\tiny $\blacksquare$}}
\def\qed{\relax\ifmmode\hskip2em \Box\else\unskip\nobreak\hskip1em $\Box$\fi}

\newtheorem{theorem}{Theorem}
\newtheorem{itlemma}{Lemma}
\newtheorem{itdefinition}{Definition}
\newtheorem{itproposition}{Proposition}
\newtheorem{itresult}{Result}
\newtheorem{itremark}{Remark}
\newtheorem{itassumption}{Assumption}
\newtheorem{itcorollary}{Corollary}
\newtheorem{itexample}{Example}

\newenvironment{proposition}{\begin{itproposition}\rm}{\end{itproposition}}
\newenvironment{remark}{\begin{itremark}\rm}{\end{itremark}}
\newenvironment{assumption}{\begin{itassumption}\rm}{\end{itassumption}}
\newenvironment{lemma}{\begin{itlemma}\rm}{\end{itlemma}}

\AtBeginDocument{
	\addtolength{\abovedisplayskip}{-1ex}
	\addtolength{\abovedisplayshortskip}{-1ex}
	\addtolength{\belowdisplayskip}{-1ex}
	\addtolength{\belowdisplayshortskip}{-1ex}
}

\title{\LARGE Dynamic quantized consensus under DoS attacks:\\ Towards a tight zooming-out factor}

\author{
	Shuai Feng, Maopeng Ran, Hideaki Ishii, Shengyuan Xu$^*$
	\thanks{Shuai Feng and Shengyuan Xu are with the School of Automation, Nanjing University of Science and Technology, Nanjing 210094, China ({\tt\small s.feng@njust.edu.cn}, {\tt syxu@njust.edu.cn}). Maopeng Ran is with the School of Automation Science and Electrical Engineering, Beihang University, Beijing, 100191, China and and also with the Zhongguancun Laboratory, Beijing, 100094, China ({\tt \small ranmp@buaa.edu.cn}). 
 Hideaki Ishii is with the Department of Computer Science, Tokyo Institute of Technology, Yokohama 226-8502, Japan ({\tt\small ishii@c.titech.ac.jp}). Corresponding author: Shengyuan Xu.}
}

\begin{document}
\maketitle
\begin{abstract}
This paper deals with dynamic quantized consensus of dynamical agents in a general form under packet losses induced by Denial-of-Service (DoS) attacks.  
The communication channel has limited bandwidth and hence the transmitted signals over the network are subject to quantization. 
To deal with agent's output, an observer is implemented at each node. The state of the observer is quantized by a finite-level quantizer and then transmitted over the network. 
To solve the problem of quantizer overflow under malicious packet losses, a zooming-in and out dynamic quantization mechanism is designed. By the new quantized controller proposed in the paper, the zooming-out factor is lower bounded by the spectral radius of the agent's dynamic matrix. 
A sufficient condition of quantization range is provided under which the finite-level quantizer is free of overflow. A sufficient condition of tolerable DoS attacks for achieving consensus is also provided. 
At last, we study scalar dynamical agents as a special case and further tighten the zooming-out factor to a value smaller than the agent's dynamic parameter. Under such a zooming-out factor, it is possible to recover the level of tolerable DoS attacks to that of unquantized consensus, and the quantizer is free of overflow.  
\end{abstract}


\vspace{-3mm}

\section{Introduction}

Consensus of multi-agent systems has a variety of applications such as distributed computation, collaborative surveillance, sensor fusion and vehicle platoon \cite{FB-LNS}. Sophisticated devices such sensors, micro computers and wireless communication blocks are embedded in each agent, which can significantly promote the performance of consensus and make cooperation among agents possible.  
However, the challenges of malicious cyber attacks causing malfunctions also emerge, such as deceptive attacks and Denial-of-Service (DoS) attacks \cite{cheng2017guest, teixeira2015secure}. Deceptive attacks influence the integrity of data and DoS attacks can induce malicious packet drops by radio-frequency interference and/or flooding the target with an overwhelming flux of packets to name a few.

In this paper, we are interested in a consensus problem of multi-agent systems under quantized data and malicious packet dropouts induced by DoS attacks. The quantizer has finite levels and quantization range. By controller design, we need to ensure that the quantizer should not saturate. Such problem is not trivial because, for example, if a dynamical agent's measurements drift/diverge during packet-dropping intervals, some measurements may exceed the range of the quantizer. 

The problems of packet dropouts have been well studied in the last two decades, e.g. in \cite{you2010minimum} for stochastic packet losses, and the case of DoS attacks inducing malicious packet dropouts has been investigated in \cite{de2015input, lu2017input, feng2020tac, li2017sinr, Deng}. 
For multi-agent systems under malicious packet dropouts, there are some recent developments for consensus, output regulation and formation control \cite{senejohnny2017jamming, feng2019secure, Deng, tang2020event}. 
Single integrator multi-agent systems under DoS attacks were studied in the papers \cite{senejohnny2017jamming, cetinkaya2020randomized}. The objective therein is practical state consensus. Specifically, in \cite{senejohnny2017jamming}, by self-triggered control, the nodes can achieve practical consensus if the number of consecutive packet losses is finite. In \cite{cetinkaya2020randomized}, by proposing randomized transmissions, the nodes are shown to achieve practical consensus even under frequent DoS attacks. For dynamical agents in a general form, the work \cite{feng2019secure} is one of the early papers considering consensus problems under DoS attacks by event-triggered control. 
The authors in \cite{deng2022resilient} propose a novel distributed resilient control method without the pre-knowledge of the leader to solve the practical cooperative output regulation problem for multi-agent systems under DoS attacks.
In \cite{tang2020event}, formation control of nonlinear multi-agent systems in the presence of malicious packet dropouts induced by DoS attacks is investigated.
In \cite{lu2018distributed}, the authors consider asynchronous DoS attacks, i.e., the attacker is able to launch independent DoS attacks on communication edges.


Wireless communication for data transmission is widely used in cyber-physical systems. Despite the advantages of wireless communication such as remote transmission and lower costs for mass devices, the ``competition" among devices for bandwidth resources may become overwhelming. Such competition would induce problems (e.g., delay and packet losses) to those control systems with large amounts of data exchange, e.g., large-scale multi-agent systems and sensor networks. 
Under limited bandwidth, signals can be subject to coarse quantization, and the consequence of quantization on control/measurement signals needs to be taken into account at the design stage. 
Static and dynamic quantizations have been proposed for various control problems. Centralized control systems under quantized communication have been extensively studied in the last two decades, for example by the seminal papers \cite{1310461, nair2004stabilizability}. Dynamic quantization with zooming-in and out is necessary to ensure quantizer unsaturation and asymptotic stability \cite{liberzon2007input}, where in particular the zooming-out factor should be able to compensate the influence of state divergence during open-loop control intervals, e.g., it should be ``larger" than the unstable eigenvalues in case of a discrete-time system \cite{you2010minimum}. Centralized systems under dynamic quantization and DoS-induced packet losses have been studied in \cite{feng2020tac, liu2021resilient}. In general, it is difficult to make the zooming-out factor equal to or smaller than the unstable eigenvalues because the state or estimation error can diverge at the rate of the unstable eigenvalues under open-loop mode (due to packet losses or sampled data control). 

Dynamic quantization has also been extended to multi-agent systems without packet losses \cite{you2011network, li2010distributed, 9429920, qiu2015quantized}. In such works, quantized systems are equipped with zooming-in capability to ensure finite data-rate quantization and asymptotic convergence. When multi-agent systems are subject to DoS attacks and limited data-rate quantization, the recent paper \cite{feng2020arxiv} provides a design of dynamic quantization with both zooming-in and out capabilities to ensure quantizer unsaturation and consensus.

For multi-agent systems, when dynamic quantization due to limited data rate and packet losses coexist, one may have a question: Can we make the zooming-out factor tight and directly link it to the agent dynamics, e.g., lower bounded by the unstable eigenvalues of the open-loop system dynamic matrix? As mentioned before, such a problem has been well addressed for centralized systems \cite{1310461, you2010minimum}, but it is still open for multi-agent systems. Moreover, a tight zooming-out factor is meaningful because it can promote the multi-agent systems' resilience against DoS attacks \cite{feng2020arxiv,9954904}.

In this paper, we provide a new design of dynamic quantized controller for output feedback multi-agent systems under DoS attacks. 
In the following, we compare our paper with the relevant literature in order to clarify our contributions. Generally speaking, the most relevant paper is our previous work \cite{feng2020arxiv}. In \cite{feng2020arxiv}, the connections of zooming-out factor to the spectral radius of the open-loop system dynamic matrix of the agent are not explicit during the computation, and its value is conservative. It is also computationally intense in the sense that it needs matrix calculations of ``high" dimensions (i.e. multi-agent's state dimension times the total number of agents) for as many rounds as the number of the maximum consecutive packet losses. This also implies that one needs to know that number before controller design. By contrast, by the new control design in this paper, to obtain the zooming-out parameter, one only needs to calculate the spectral radius of the agent system matrix, and chooses a value larger than it. The information about the maximum DoS-induced packet losses is not needed. Moreover, an observer is implemented in this paper for handling output feedback while \cite{feng2020arxiv} does not involve state observation. 
From the viewpoint of technical analysis by switched system theory, this paper is more involved for having four modes, while \cite{feng2020arxiv} has only two modes.    
Compared with \cite{senejohnny2017jamming}, our paper additionally takes limited data rate and the induced quantizer overflow problem into consideration. Moreover, the model of multi-agent systems in \cite{senejohnny2017jamming} takes the form of a single integrator, in which consensus error does not diverge even in the presence of DoS attacks. In our paper, the model of multi-agent systems is more general, which can incorporate open-loop unstable dynamics. Technically, it is more challenging to achieve consensus for multi-agent systems with unstable dynamics under DoS, not to mention under finite-level quantization. Compared with \cite{feng2019secure}, our paper additionally considers quantized control under DoS, though transmissions are periodic and not based on event-triggered control.


In view of the comparisons mentioned above, we summarize the main contributions of this paper:

a) We develop a new \emph{output feedback} quantized controller design for multi-agent systems, in which the bound of zooming-out factor is tight and directly linked to the agent dynamics compared with \cite{feng2020arxiv}: it is \emph{lower bounded by the spectral radius} of the agent's system matrix. Sufficient conditions on quantization range for overflow prevention and DoS attack for consensus are provided.     

b) For scalar multi-agent systems, beyond the results of general linear multi-agent systems in a), we provide an approach to find a \emph{tighter} zooming-out factor \emph{smaller} than the agent's system dynamic parameter. Under such a zooming-out factor, the bound of tolerable DoS attacks under unquantized consensus is recovered, and the quantizer is not saturated.

This paper is organized as follows. In Section II, we introduce the framework consisting of multi-agent systems and the class of DoS attacks. Section III contains two parts: general multi-agent systems as the main part and scalar multi-agent systems as a special case. 
A numerical example is presented in Section IV, and finally Section V ends the paper with concluding remarks and future research.

\textbf{Notation.} We denote by  $\mathbb R$ the set of reals. Given $b \in \mathbb R$, $\mathbb R_{\geq b}$ and $\mathbb R_{>b}$ denote the sets of reals no smaller than $b$ and reals greater than $b$, respectively; $\mathbb R_{\le b}$ and $\mathbb R_{<b}$ represent the sets of reals no larger than $b$ and reals smaller than $b$, respectively; $\mathbb Z$ denotes the set of integers. For any $c \in \mathbb Z$, we denote $\mathbb Z_{\ge c} := \{c,c+ 1,\cdots\}$. 
Given a vector $y$ and a matrix $\Gamma$, let $\|y\|$ and $\|y\|_\infty$ denote the $2 $- and $\infty$-norms of vector $y$, respectively,  
and $\|\Gamma\|$ and $\|\Gamma\|_\infty$ represent the corresponding induced norms of matrix $\Gamma$. Moreover, $\rho(\Gamma)$ denotes the spectral radius of $\Gamma$. Given an interval $\mathcal{I}$, $|\mathcal{I}|$ denotes its length. The Kronecker product is denoted by $\otimes$.

\section{Framework}
\textbf{Communication graph.}
We let graph $\mathcal{G} = (\mathcal{N},\mathcal{E})$ denote the communication topology among $N$  agents, where $\mathcal{N}=\{1, 2, \cdots, N \}$ denotes the set of agents and $\mathcal{E} \subseteq \mathcal{N} \times \mathcal{N}$ denotes the set of edges. Let $\mathcal N_i$ denote the set of the neighbors of agent $i$, where $i=1, 2, \cdots, N$. In this paper, we assume that the graph $\mathcal{G}$ is undirected and connected, i.e. if $j \in \mathcal N_i$, then $i \in \mathcal N_j$. Let $A _ {\mathcal G}= [a_{ij}] \in \mathbb{R}^{N\times N} $ denote the adjacency matrix of the graph $\mathcal {G}$, where $a_{ij} > 0$ if and only if $j \in \mathcal N_i$ and $a_{ii}=0$. Define the Laplacian matrix $L_ \mathcal{G} = [l_{ij}]  \in \mathbb{R}^{N\times N} $, in which $l_{ii} = \sum_{j = 1 }^{N} a_{ij}$ and $l_{ij} = - a_{ij} $ if $i \ne j $. Let $\lambda_i$ ($i=1, 2, \cdots, N$) denote the eigenvalues of $L_ \mathcal{G}$ and in particular we have $\lambda_1 = 0$ due to the graph being connected. Let $I_N$ denote an identity matrix with dimension $N$.

\subsection{System description}

The agents interacting over the network $\mathcal{G}$ are homogeneous linear multi-agent systems with sampling period $\Delta \in \mathbb{R}_{>0} $:
\begin{subequations}\label{system}
\begin{align}
x_i(k\Delta) &= A x_i((k-1)\Delta) +B u_i((k-1)\Delta) \\
y_i(k\Delta)&= C x_i(k\Delta)
\end{align}
\end{subequations}
where $i \in \mathcal N$, $k\in \mathbb{Z}_{\ge 1}$, $A\in \mathbb R^{n \times n}$, $B\in \mathbb R^{n \times w}$ and $C\in \mathbb R^{v\times n}$. We assume that $(A,B)$ is stabilizable and $(A,C)$ is  observable. 
$x_i(k \Delta)\in \mathbb{R}^{n}$ denotes the state of agent $i$ and $y_i(k \Delta)\in \mathbb{R}^{v}$ denotes the output. We assume that an upper bound of the initial condition $x_i(0) \in  \mathbb{R}^{n}$ is known, i.e. $\| x_i(0)\|_\infty \le C_{x_0} \in \mathbb{R}_{>0}$ \cite{you2011network, li2010distributed, 9429920}. 
Here, $C_{x_0}$ can be an arbitrarily large real as long as it satisfies this bound. This is for preventing the overflow of state quantization at the beginning.
Let $u_i((k-1)\Delta) \in \mathbb{R}^{w}$ denote the control input, whose computation will be given in (\ref{controller0}).

We introduce the control objective of this paper: State consensus. The average of the states is computed as
$
\overline{x}(k\Delta) = \frac{1}{N}\sum_{i=1}^N x_i(k\Delta) 
$
and state consensus is defined by 
\begin{align}\label{control objective}
\lim_{k  \to \infty }   x_i(k\Delta) - \overline  x(k\Delta)    = 0, \,\,\, i=1,2,\cdots, N.
\end{align}
It is trivial that (\ref{control objective}) also implies the consensus of output $y_i(k\Delta)$. We assume that the spectral radius $\rho(A) \ge 1$. Otherwise, consensus (\ref{control objective}) is trivially achieved by setting $u_i(k\Delta)=0$ for all $k$.

In this paper, the communication channel for information exchange is bandwidth limited and subject to DoS. 
We assume that the transmission attempts of each agent take place periodically at time $k\Delta$ with $k\in \mathbb{Z}_{\ge 1}$ and free of delay. 
Agent $i = 1, 2, \cdots, N$ can only exchange quantized information with its neighbor agents $j \in \mathcal N_i$ over the network $\mathcal G$ due to bandwidth constraints. In the presence of DoS, transmission attempts fail. We let $\{s_r\} \subseteq \{k\Delta\}$ represent the instants of successful transmissions. Note that $s_0 \in \mathbb{R}_{\ge \Delta}$ is the instant when the first successful transmission occurs. Also, we let $s_{-1}$ denote the time instant $0$. 
In the remainder of the paper, we 
let $k$ represent instant $k\Delta$, and $s_r + p$ represent instant $s_r + p\Delta$ ($p \in \mathbb Z _{\ge 0}$) for the ease of notation.


\textbf{Uniform quantizer.} Let $\chi \in \mathbb{R}$ be the original scalar value before quantization and $q_R(\cdot)$ be the quantization function for scalar
input values as
\begin{align}\setlength{\arraycolsep}{3pt}  \label{quantizer}
q_R (\chi) = 
\left\{
\begin{array}{lll}
0 & -\sigma < \chi < \sigma & \\
2z \sigma & (2z-1)\sigma \le \chi < (2z+1)\sigma  \\
2R\sigma & \chi \ge  (2R+1) \sigma&         \\
-q_R (-\chi) & \chi \le - \sigma & 
\end{array}
\right.
\end{align}
where $R\in \mathbb{Z}_{>0}$ is to be designed and $z =1, 2, \cdots, R$, and $\sigma \in \mathbb{R}_{>0}$. If the quantizer is unsaturated such that $|\chi| \le (2R+1)\sigma $, then the error induced by quantization satisfies 
\begin{align}\label{quantizer error}
|\chi - q_R(\chi)| \le \sigma, \,\, \text{if}\,\, |\chi| \le (2R+1)\sigma.
\end{align}
 Moreover, we define the vector version of the quantization function as $Q_R(\beta) = [\,q_R(\beta_1)\,\,q_R(\beta_2)\,\, \cdots \,\, q_R(\beta_f) \, ]^T \in \mathbb R ^f$, where $\beta = [\beta_1\,\, \beta_ 2\,\, \cdots \beta_f]^ T \in \mathbb R ^f$ with $f \in \mathbb Z_{\ge 1} $. 
 It is clear that $\beta$ can be properly quantized if $\|\beta\|_\infty \le (2R+1)\sigma$.  
In the remainder of the paper, by quantizer overflow or saturation, we mean that at least a $\beta_i$ exceeds the range of quantization, i.e., $\|\beta\|_\infty > (2R+1)\sigma$ or equivalently $|\beta_i | > (2R+1)\sigma$.

\vspace{-2mm}

\subsection{Time-constrained DoS}
\vspace{-1mm}
In this paper, we refer to DoS as the event for which all the encoded signals cannot be received by the decoders and it affects all the agents.  
We consider a general DoS model developed in \cite{de2015input}
that describes the attacker's action by the frequency of DoS attacks and their duration. Let 
$\{h_q\}_{q \in \mathbb Z_0}$ with $h_0 \geq \Delta$ denote the sequence 
of DoS \emph{off/on} transitions, that is,
the time instants at which DoS exhibits 
a transition from zero (transmissions are successful) to one 
(transmissions are not successful).
Hence,
$
H_q :=\{h_q\} \cup [h_q,h_q+\tau_q[  
$
represents the $q$-th DoS time-interval, of a length $\tau_q \in \mathbb R_{\geq 0}$,
over which the network is in DoS status. If $\tau_q=0$, then
$H_q$ takes the form of a single pulse at $h_q$.  
Given $\tau,t \in \mathbb R_{\geq0}$ with $t\geq\tau$, 
let $n(\tau,t)$
denote the number of DoS \emph{off/on} transitions
over $[\tau,t]$, and let 
$
\Xi(\tau,t) := \bigcup_{q \in \mathbb Z_0} H_q  \, \cap  \, [\tau,t] 
$
be the subset of $[\tau,t]$ where the network is in DoS status. 

\begin{assumption}
(\emph{DoS frequency}) \cite{de2015input}. 
	There exist constants 
	$\eta \in \mathbb R_{\geq 0}$ and 
	$\tau_D \in \mathbb R_{> 0}$ such that
$
	n(\tau,t)  \, \leq \,  \eta + \frac{t-\tau}{\tau_D}
$
	for all  $\tau,t \in \mathbb R_{\geq \Delta}$ with $t\geq\tau$.
	\qedp
\end{assumption}

\begin{assumption} 
	(\emph{DoS duration}) \cite{de2015input}. 
	There exist constants $\kappa \in \mathbb R_{\geq 0}$ and $T  \in \mathbb R_{>1}$ such that
$
	|\Xi(\tau,t)|  \, \leq \,  \kappa + \frac{t-\tau}{T}
$
	for all  $\tau,t \in \mathbb R_{\geq \Delta}$ with $t\geq\tau$. 
	\qedp
\end{assumption}

\begin{remark}
	Assumptions 1 and 2
	do only constrain a given DoS signal in terms of its average frequency and duration.
	By \cite{hespanha1999stability}, $\tau_D$ can be considered as the average dwell-time between 
	consecutive DoS off/on transitions, while $\eta$ is the chattering bound.
	Assumption 2 expresses a similar 
	requirement with respect to the duration of DoS. 
	It expresses the property that, on the average,
	the total duration over which communication is 
	interrupted does not exceed a certain \emph{fraction} of time specified by $1/T$.
	Like $\eta$, the constant $\kappa$ is a regularization term. It is needed because
	during a DoS interval, one has $|\Xi(h_q,h_q+\tau_q)| = \tau_q >  \tau_q /T$.
	Thus $\kappa$ serves to make Assumption 2 consistent. 
	Conditions $\tau_D>0$ and $T>1$ imply that DoS cannot occur at an infinitely
	fast rate and be always active. \qedp
\end{remark}

The lemma below presents the relation between the number of successful transmissions and the number of transmission attempts $k$. Let $T_S(1,k)$ and $T_U(1,k)$ denote the numbers of successful and unsuccessful transmissions between steps 1 and $k$, respectively. 

\begin{lemma} \label{Lemma T} 
	\cite{feng2020tac} Consider the DoS attacks in Assumptions 1 and 2 and the network sampling period $\Delta$. If 
	$1/T+ \Delta /\tau_D <1$, then 
	$
	T_S(1, k)
	\ge  \left(1- 1/T - \Delta/\tau_D \right) k   - \frac{\kappa+\eta\Delta}{\Delta}.  
	$ \qedp
\end{lemma}

\vspace{-2mm}
\section{Main results}
\vspace{-1mm}

\subsection{Control architecture}
\vspace{-1mm}
To deal with output feedback, an observer estimating $ x_i (k)$ from $y_i(k)$ is locally implemented at agent $i \in \mathcal N$ given by
\begin{align}\label{local observer}
\hat x_i(k)&=A \hat x_i(k-1)  + B u_i(k-1)  \nonumber\\
&\,\,\,+ F(y_i(k-1)-C \hat x_i (k-1))
\end{align}
where $\hat x_i(k) \in \mathbb{R} ^ n $ denotes the observer state and $\hat x_i(0)=0$.
Since $(A,C)$ is observable, there exists a matrix $F\in \mathbb R ^{n\times v}$ such that the spectral radius $\rho(A-FC)<1$.   
Then, $\hat x_i(k)$ is quantized and transmitted to the neighbors. Let $\tilde x _ i ^ j(k) \in \mathbb{R} ^ n $ denote the decoded value of $\hat x_i(t)$ by agent $j \in \mathcal N_i$ and vice versa for $\tilde x _ j ^ i(k)$, whose computations will be given later.

For agent $i\in \mathcal N$, the control input $u_i(k)$ has two modes aligned with the status of DoS: 
\begin{align}\label{controller0}
u_i(k) = 
\left\{
\begin{array}{ll}
K \sum_{j=1} ^{N} a_{ij} (\tilde x_j ^ i (k) - \tilde x_i ^ i (k)) &\text{if}\,\,k \notin H_q \\
0 & \text{if}\,\, k \in H_q
\end{array}
\right.
\end{align}
for $k=1, 2, \cdots$. Let $u_i(0)=0$. Note that each agent is able to passively know the status of DoS at $k$ by whether receiving neighbors' transmissions or not. 
We assume that there exists a feedback gain $K \in \mathbb R ^{w \times n}$ such that the spectral radius of
\begin{align}
J= \text{diag}(J_2,  \cdots, J_N), J_i=A- \lambda_i BK,\,\,\, i=2,...,N 
\end{align}
satisfies $\rho(J) < 1$, where $\lambda_i$ denotes the eigenvalues of $L _{\mathcal G}$. Note that $\rho(J) < 1$ is needed to achieve consensus even if DoS is absent. The assumption on the existence of $K$ is motivated by the consensusability of linear discrete-time multi-agent systems \cite{you2011network}. If DoS is absent and precise information is available, a sufficient condition to ensure the existence of $K$ is that
$
		(A,B) ~\textrm{stabilizable and} ~\prod_{p}|\lambda_p^u(A)|<\frac{1+\lambda_2/\lambda_N}{1-\lambda_2/\lambda_N}, 
$
	where $\lambda_p^u(A)$ represents unstable eigenvalues of $A$.

In our paper, the encoder and decoder for each state have the same structure. In the following, we first explain the decoding process. 
In (\ref{controller0}), $\tilde{x}_j^i(k)$ is the decoded value of $\hat x_j(k)$ ($j\in \mathcal N_i$) at the decoder of agent $i$ with initial condition $\tilde{x}_j^i(0)$. Its computation is given by
\begin{eqnarray}\setlength{\arraycolsep}{3pt}  \label{eq estimator i}
	\tilde x_j ^ i(k)
	=
	\left\{
	\begin{array}{ll}
		A  \tilde x_j ^ i(k-1) + \theta(k-1) \hat Q_j ^ i (k)  & \text{if $k \notin H_q$ } \\
		A\tilde x_j ^ i (k-1)               & \text{if $k \in  H_q$} 
	\end{array}
	\right.
\end{eqnarray}
in which $k=1, 2, \cdots$, and $\hat{Q}_j^i(k)$ is the value generated and transmitted by the encoder of agent $j$ given by 
\begin{eqnarray}\label{Q_i i}
	\hat  Q_j ^i (k) = Q_R \left(\frac{\hat x_j(k) -  A \tilde x_j ^ i (k-1)}{\theta(k-1)}  \right).
\end{eqnarray} 
The computation of $\tilde x _ i ^ i(k) $ in the decoder of agent $i$ follows
\begin{align}\setlength{\arraycolsep}{3pt}  
&	\tilde x_i ^ i(k)
	\!=\!
	\left\{
	\begin{array}{ll}
		A  \tilde x_i ^ i(k-1) + \theta(k-1) \hat Q_i ^ i (k)  & \text{if $k \notin H_q$ } \\
		A\tilde x_i ^ i (k-1)               & \text{if $k \in  H_q$} 
	\end{array}
	\right. \label{10}  \\
& 	\hat  Q_i ^i (k) = Q_R \left(\frac{\hat x_i(k) -  A \tilde x_i ^ i (k-1)}{\theta(k-1)}  \right). \label{11} 
\end{align}

Now we explain the synchronization of a decoded state among agents, that is, for $i, l \in \mathcal N _ j$, the values of $\tilde x_j ^ i (k)$, $\tilde x_j ^ l (k)$ and $\tilde x_j ^ j (k)$ are identical for all $k$. This is because the decoders of agents $i, j$ and $l$ have the same initial condition ($\tilde x_j ^ i (0)=\tilde x_j ^ l (0)=\tilde x_j ^ j (0)=0$), have the same $\theta(k)$ and receive the same $Q_R ( (\hat x_j(k) -  A \tilde x_j ^ {i, l, j} (k-1))/\theta(k-1) )$ in $\hat Q_{j} ^ i (k)$,  $\hat Q_{j} ^ l (k)$ and $\hat Q_{j} ^ j (k)$ for all $k$. The synchronization of $\theta(k)$ in the decoders and also in the encoders will be explained after (\ref{eq h}). Thus, the superscripts of $\tilde x_j ^ i (k)$, $\tilde x_j ^ l (k)$ and $\tilde x_j ^ j (k)$  can be omitted, and it is enough to let $\tilde x_j(k)$ represent the decoded value of $\hat x_j (k)$ in the decoders. At last, we point out that the encoder is a copy of the decoder, and thus one has that $\tilde x_j (k)$ in the encoder also has the same value as in the decoders.

Therefore (\ref{controller0}) can be rewritten as
\begin{align}\label{controller}
u_i(k) = 
\left\{
\begin{array}{ll}
K \sum_{j=1} ^{N} a_{ij} (\tilde x_j   (k) - \tilde x_i   (k)) &\text{if}\,k \notin H_q \\
0 & \text{if}\, k \in H_q
\end{array}
\right.
\end{align}
and (\ref{eq estimator i}) and (\ref{10}) can be rewritten as
\begin{eqnarray}\setlength{\arraycolsep}{3pt}  \label{eq estimator}
\tilde x_j (k)
=
\left\{
\begin{array}{ll}
A  \tilde x_j(k-1) + \theta(k-1) \hat Q_j(k)  & \text{if $k \notin H_q$ } \\
A\tilde x_j(k-1)               & \text{if $k \in  H_q$} 
\end{array}
\right.
\end{eqnarray}
in which $k=1, 2, \cdots$, $j\in\{i\}\cup\mathcal{N}_i$ with $\tilde x_j (0)=    0$. Similarly, (\ref{Q_i i}) and (\ref{11}) can be written as 
\begin{eqnarray}\label{Q_i} 
\hat  Q_j (k) = Q_R \left(\frac{\hat x_j(k) -  A \tilde x_j(k-1)}{\theta(k-1)}  \right), \,\, k=1, 2, \cdots.
\end{eqnarray}
The switched-type estimator in (\ref{eq estimator}) has the following motivations. The first equation in (\ref{eq estimator}) is for acquiring the quantized information of $\hat x_j (k)$ under successful transmissions, and then calculates $\tilde x_j (k)$. This step is also necessary for the DoS-free case \cite{you2011network}. The second equation in (\ref{eq estimator}) is an open loop estimation, and it together with the second equation in (\ref{controller}) can decouple the state of the agents and the quantization errors (see Case d) later).

A key parameter in (\ref{Q_i}) is the scaling parameter $\theta (k-1)$. By adjusting its value dynamically, the variable to be quantized will be kept within the bounded quantization range without saturation. 
The scaling parameter $\theta(k) \in \mathbb{R}_{>0}$ is updated as
\begin{align}\label{eq h}
\theta(k) = 
\left\{
\begin{array}{ll}
\gamma_1 \theta(k-1) & \quad \text{if $k  \notin H_q $} \\
\gamma_2 \theta(k-1) & \quad \text{if $k  \in H_q $  }
\end{array}
\right.  \,\,\, k=1, 2, \cdots
\end{align}
with $\theta(0)  = \theta_0 \in \mathbb{R}_{>0}$. 
The parameters $\gamma_1$ and $\gamma_2$ are the so-called zooming-in and out factors in dynamic quantization, respectively. 
Since $\theta(k)$ in the encoders and decoders has the same initial condition $\theta(0)$, and $k \in H_q$ or $k \notin H_q$ is passively known as mentioned before, one can see that $\theta(k)$ is synchronized in all the encoders and decoders. 
The zooming-in and out mechanism in (\ref{eq h}) is majorly inspired by \cite{you2010minimum, feng2020tac} studying centralized systems. The scaling parameter is a dynamic sequence whose increasing and decreasing depend on DoS attacks in order to mitigate the influence of DoS-induced packet losses. 
If one assumes that each agent is aware of its own $J_i$ and their neighbors' $J_j$ ($j \in \mathcal N_i$), it is possible to design distributed zooming-in factors, namely $\gamma_1 ^ i$. Then the new $\theta(k)$ and $\gamma_1$ are complicated and will be a vector and a matrix, respectively. This case will be left for future research. The design of zooming-out factor may not change since it is dependent on $A$ (see Lemma \ref{Lemma3} later).

During DoS intervals, $\hat x_j(k) -  A \tilde x_j(k-1)$ in (\ref{Q_i}) may diverge. Therefore, the scaling parameter $\theta(k-1)$ must increase using $\gamma_2$ so that $Q_R(\cdot)$ is not saturated. If the transmissions succeed, the quantizers zoom in and $\theta(k)$ decreases by using $\gamma_1$.   
The selections of $\gamma_1$ and $\gamma_2$ will be specified later, and one of the objectives of this paper is to find a $\gamma_2$ as tight as possible. If one assumes that the communication network is free from any DoS-induced or random packet losses, then the zooming-out mechanism is not necessary since $\hat x_j(k) -  A \tilde x_j(k-1)$ in $Q_R(\cdot)$ does not diverge. In this case, one only needs the zooming-in mechanism to ensure the property of asymptotic convergence \cite{you2011network, 9429920}.

By defining vectors $\tilde x(k) = [\tilde  x_1 ^T (k) \cdots \tilde  x_N ^T (k)]^T, \hat x(k) = [\hat x_1 ^T  (k)  \cdots \hat x_N ^T (k)]^T, Q(k) = [\hat Q_1 ^T (k)  \cdots \hat Q_N ^T  (k)]^T  \in \mathbb{R}^{nN}$, one can obtain the compact form of (\ref{eq estimator}) as
\begin{eqnarray}\setlength{\arraycolsep}{3pt}  \label{eq estimator all}
\tilde x(k)  
=
\left\{
\begin{array}{ll}
A_N\tilde x(k-1) + \theta(k-1)Q(k) & \text{if $k \notin H_q$} \\
A_N \tilde x(k-1)        & \text{if $k \in H_q$} 
\end{array}
\right.
\end{eqnarray}
for $k=1, 2, \cdots$. The compact form of the observer in (\ref{local observer}) is 
\begin{align}\label{compact observer}
\hat x(k)=&A_N \hat x(k-1)  + B_N u(k-1)  \nonumber\\
&+ F_N(y(k-1)-C_N \hat x (k-1)) 
\end{align}
in which $A_N:=I_N \otimes A$, $B_N:=I_N \otimes B$, $C_N:=I_N \otimes C$, $F_N:=I_N \otimes F$, $u(k)=[u_1 ^T(k) \cdots u_N^T(k)]^T \in \mathbb R ^{N  w}$ and $y(k)=[y_1 ^T(k) \cdots y_N^T(k)]^T \in \mathbb R ^{N   v} $.
Let $x(k) = [x_1 ^T (k)\cdots x_N ^T (k)]^T \in \mathbb{R}^{nN}$. 
With the vectors $x(k), \tilde x(k)$ and $\hat x(k)$, we define the following errors in vector form
\begin{align*}
&\!\!\!\!  e_c(k)\!=\!\hat x(k) \!   -\! \tilde x(k) \!   =\! [\hat x_1 ^T(k) \!-\! \tilde x_1 ^T (k) \!\cdots\! \hat x_N ^T (k) \!-\! \tilde x_N  ^T(k)]^T  \\
&\!\!\!\!e_o(k) \!=\! x(k) \!   - \! \hat  x(k) \!=\! [ x_1  ^T (k) \!-\! \hat x_1  ^T (k)\!\cdots\! x_N  ^T (k) \!-\! \hat x_N  ^T (k)]^T
\end{align*}
in which $e_c(k)$ denotes the error due to signal coding and $e_o(k)$ denotes the observer error.
From (\ref{system}), one can see that the update of $x(k)$ is independent on $k \in H_q$ or $k \notin H_q$. However, the update of $x(k)$ depends on $k-1 \in H_q$ or $k-1 \notin H_q$, due to the control law $u_i(k-1)$ in (\ref{controller}). Note that such dependence does not exist in \cite{feng2020arxiv}.
One can obtain the compact form of the closed-loop dynamics of $x(k)$ as
\begin{align}\label{eq process all} 
&x(k) = \nonumber\\
&\left\{\!\!\!\!
\begin{array}{ll}\setlength{\arraycolsep}{0.1pt} 
 G x(k-1) \! + \! L(e_o(k-1) \!+\! e_c(k-1)) & \!\!\!\! \text{if}\, k\!-\!1 \!\notin\! H_q \\
A_N x(k-1) & \!\!\!\! \text{if}\, k\!-\!1 \!\in\! H_q 
\end{array}
 \right.
\end{align}
where $G:=A_N - L_\mathcal G \otimes BK$ and 
$
L := L_ \mathcal{G} \otimes BK.
$
Let the discrepancy between the state of agent $i$ and $\overline{x}(k)$ be $\delta_i(k) : = x_i(k) - \overline x(k) \in \mathbb{R} ^n$ and let $\delta(k) = [\delta_1 ^T  (k) \,\,\cdots \,\, \delta_N  ^T (k)]^T \in \mathbb{R}^{nN}$. Then one can obtain the compact dynamics of $\delta(k)$:
\begin{align}\label{delta all}
&\delta(k) = \nonumber\\
&\left\{\!\!\!\!
\begin{array}{ll}\setlength{\arraycolsep}{0.1pt} 
G \delta(k-1) \! + \! L(e_o(k-1) \!+\! e_c(k-1)) & \!\!\!\! \text{if}\, k\!-\!1 \!\notin\! H_q \\
A_N \delta(k-1) & \!\!\!\! \text{if}\, k\!-\!1 \!\in\! H_q. 
\end{array}
\right.\!\!\!\!
\end{align}
It is clear that if $\delta (k) \to 0$ as $k \to \infty$, consensus of the multi-agent system is achieved as in (\ref{control objective}).
If the multi-agent system is subject to DoS, $\delta(k)$ has a diverging mode by the second equation in (\ref{delta all}), and the dynamics of $e_c(k)$ and $e_o(k)$ are not clear, which implies that consensus may not be achieved.

Note that the control system in this paper looks similar but different from the common ``to zero" static controller, i.e., the control input $u_i(k)$ is set to zero under DoS attacks, 
see \cite{senejohnny2017jamming}. The controller design in our paper is also different from the ``pure" prediction based controller in lossy networks (i.e., $u_i(k)$ is always computed based on the state prediction) in \cite{feng2020arxiv, feng2019secure}. In our paper, the decoder (\ref{eq estimator}) generates $\hat x_j(k)$ ($j \in \{i\}\cup\mathcal N_i  $) regardless of the presence of DoS. Meanwhile, the input $u_i(k)$ is set to zero and does not depend on the estimated state $\hat x_j(k)$ when $k \in H_q$, though $\hat x_j(k)$ is available. However, computing $\hat x_j(k)$ under DoS is necessary since its value is useful when DoS is over. Consequently, as will be shown later, the technical analysis involves four modes by expressing the overall system as a switched system. By contrast, the analysis in \cite{feng2020arxiv} needs to consider only two modes.

\subsection{Dynamics of the multi-agent systems}
Though the update of $\delta(k)$ in (\ref{delta all}) depends on the previous step, i.e., $k-1  \notin H_q$ or $k-1 \in H_q$, the update of $e_c(k)$ is affected by both the $k-1$ and $k$-th steps. 
This is because $x(k)$ in (\ref{eq process all}) depends on $k-1 \notin H_q $ or $k-1 \in H_q$, and $\hat x(k)$ in (\ref{eq estimator all}) depends on $k \notin H_q $ or $k\in H_q$. To make the technical analysis approachable, we conduct the analysis by four cases:
\begin{align*}
\begin{array}{ll}
	\text{a})\, k \notin H_q$ and $k-1 \notin H_q & \text{b})\, k \notin H_q$ and $k-1 \in H_q \\
\text{c})\,	k \in H_q$ and $k-1 \notin H_q &  \text{d})\, k \in H_q$ and $k-1 \in H_q.
\end{array}
\end{align*}

\textbf{Case a)} In view of $\tilde x(k)$ in the first equation of (\ref{eq estimator all}) and $\hat x(k)$ in (\ref{compact observer}), one can obtain 
$
e_c(k) = \hat x(k) - A_N \tilde x(k-1)  
- \theta(k-1) Q_R \left(\frac{ \hat x(k) - A_N \tilde x(k-1) }{\theta(k-1)}\right)
$,
in which we have
\begin{align}\label{HK}
& \hat x(k) -A_N \tilde x(k-1)    \nonumber\\
&=A_N e_c(k-1) + B_N u(k-1) + F_N C_N e_o(k-1)\nonumber\\
&= H e_c(k-1) - L\delta(k-1) +Pe_o(k-1)
\end{align}
with 
$
H := A_N +  L_ \mathcal{G} \otimes B K 
$ and $P:=L-F_NC_N$.
One can also compute the estimation error in the observer 
\begin{align}\label{observer error}
e_o(k)=(A_N - F_NC_N) e_o(k-1)
\end{align}
in which the spectral radius $\rho(A_N - F_NC_N)<1$ due to $\rho(A - FC)<1$.
Recall the update of $\delta(k)$ for $k-1 \notin H_q$ in (\ref{delta all}). Then by (\ref{delta all})--(\ref{observer error}), one obtains the dynamics of $\delta(k)$, $e_c(k)$ and $e_o(k)$ as follows: 
\begin{subequations}\label{case a}
\begin{align}
&\delta(k) =G \delta(k-1)  + L(e_o(k-1) + e_c(k-1))\\
&e_c(k) 
=\, H e_c(k-1) - L\delta(k-1)+Pe_o(k-1)   \nonumber\\ 
   &\!\!   -\!  \theta(k\!-\!1) Q_R\! \left(\!\!\frac{ H e_c(k-1) \!-\! L\delta(k-1) \!+\! Pe_o(k-1) }{\theta(k-1)} \!\!\right)\label{e no DoS}\\
 &e_o(k)=(A_N - F_NC_N) e_o(k-1). \label{e3 no dos}
\end{align}
\end{subequations}

\textbf{Case b)} In view of $\tilde  x(k)$ in the first equation in (\ref{eq estimator all}) and $\hat x(k)$ in (\ref{compact observer}), where $u(k-1)=0$ due to $k-1\in H_q$, one can obtain that
$
  e_c(k) =  A_N e_c(k-1)  +  F_NC_Ne_o(k-1)  
 - \theta(k-1)Q_R\left(\frac{ A_N e_c(k-1)  +  F_NC_Ne_o(k-1)}{\theta(k-1)} \right ).
$
The dynamics of $\delta(k)$ for Case b) follows from the second equation in (\ref{delta all}). The control input applied to the observer is also $u(k-1)=0$ due to $k-1\in H_q$, which implies that (\ref{e3 no dos}) still holds. With $ e_c(k)$ above, one obtains the dynamics of $\delta(k)$, $e_c(k)$ and $e_o(k)$ as
\begin{subequations}\label{case b}
	\begin{align}
	&\delta(k) = A_N \delta(k-1)  \\
	& e_c(k) =  A_N e_c(k-1)  +  F_NC_Ne_o(k-1) \nonumber\\
	&\,\,- \theta(k-1)Q_R\left(\frac{ A_N e_c(k-1)  + F_NC_Ne_o(k-1)}{\theta(k-1)} \right )\\
	&e_o(k)=(A_N - F_NC_N) e_o(k-1).
	\end{align}
\end{subequations}

\textbf{Case c)} By substituting the dynamics of $\tilde x(k)$ in the second equation of (\ref{eq estimator all}), one has the evolution of $e_c(k)$ as 
$
e_c(k) 
=  A_N e_c(k-1) + B_N u(k-1) + F_NC_Ne_o(k-1) 
=  H e_c(k-1) - L\delta(k-1)+Pe_o(k-1).
$
The dynamics of $e_o(k)$ in (\ref{observer error}) also holds in this case. 
In view of the dynamics of $\delta(k)$ in the first equation of (\ref{delta all}), overall one can obtain that 
\begin{subequations}\label{case c}
	\begin{align}
	\delta(k) &= G \delta(k-1) +L ( e_o(k-1) + e_c(k-1) ) \\
	e_c(k) &= H e_c(k-1) - L\delta(k-1) + Pe_o(k-1) \\
	e_o(k) &=(A_N - F_NC_N)e_o(k-1).
	\end{align}
\end{subequations}

\textbf{Case d)} For computing $e_c(k)$, by $\tilde x(k)$ in the second equation of (\ref{eq estimator all}) and $\hat x(k)$ in (\ref{compact observer}) with $u(k-1)=0$ due to $k-1 \in H_q$, one can have
$
e_c(k) 
=   A_N \hat x(k-1) +F_NC_N e_o(k-1) -A_N \tilde x(k-1) 
=  A_N  e_c(k-1) + F_NC_N e_o(k-1). 
$
The dynamics of $e_o(k)$ in (\ref{observer error}) still holds in this case. 
Then, by combining the dynamics of $\delta(k)$ in the second equation of (\ref{delta all}), we can obtain that
\begin{subequations}\label{case d}
	\begin{align}
	\delta(k) &= A_N  \delta(k-1)  \\
	e_c(k) &=  A_N  e_c(k-1) + F_NC_N e_o(k-1) \\
	 e_o(k) &= (A_N-F_NC_N) e_o(k-1).
	\end{align}
\end{subequations}

To further facilitate the analysis, we define three new variables:
\begin{align}\label{transformation}
\alpha(k) := \frac{\delta(k)}{\theta(k)},\,
\xi_c (k) := \frac{e_c(k)}{\theta(k)},\, \xi_o (k) := \frac{e_o(k)}{\theta(k)}
\end{align}
where $\theta(k)$ has been given in (\ref{eq h}). The dynamics of $\alpha(k)$, $\xi_c(k)$ and $\xi_o(k)$ corresponding to the four cases above, respectively, are presented in the following:


\textbf{Case a)} 
\begin{subequations}\label{case a 1}  
	\begin{align}
	&\alpha(k) = \frac{G}{\gamma_1} \alpha(k-1) +\frac{L}{\gamma_1}(\xi_o(k-1) + \xi_c(k-1)) \\
	&\xi_c(k) 
	=\, \frac{H}{\gamma_1} \xi_c(k-1) - \frac{L}{\gamma_1} \alpha (k-1) + \frac{P}{\gamma_1}\xi_o(k-1)   \nonumber\\ 
	&  \,\,\, -   \frac{Q_R \left(H \xi_c(k-1) - L\alpha(k-1) +  P\xi_o(k-1) \right)}{\gamma_1}\\
	&\xi_o(k) = \frac{A_N -  F_NC_N}{\gamma_1} \xi_o(k-1)
	\end{align}
\end{subequations}

\textbf{Case b)} 
\begin{subequations}\label{case b 1} 
	\begin{align}
	&\alpha(k) = \frac{A_N}{\gamma_1} \alpha(k-1)  \\
	&\xi_c(k) 
	=\,  \frac{A_N}{\gamma_1} \xi_c(k-1)  +  \frac{F_NC_N}{\gamma_1} \xi_o(k-1) \nonumber\\
	&\quad\quad \quad \,\,\,  -\frac{Q_R\left(   A_N \xi_c(k-1)  + F_NC_N \xi_o(k-1)  \right )}{\gamma_1} \\
	&\xi_o(k) = \frac{A_N - F_NC_N}{\gamma_1} \xi_o(k-1)
	\end{align}
\end{subequations}

\textbf{Case c)} 
\begin{subequations}\label{case c 1}  
	\begin{align}
	&\alpha(k) = \frac{G}{\gamma_2} \alpha(k-1) + \frac{L}{\gamma_2} (\xi_o(k-1) +\xi_c(k-1)) \\
	&\xi_c(k) =  \frac{H}{\gamma_2} \xi_c(k-1)  - \frac{L}{\gamma_2} \alpha(k-1) + \frac{P}{\gamma_2} \xi_o(k-1)\\
	&\xi_o(k)  = \frac{A_N-F_NC_N}{\gamma_2} \xi_o (k-1)
	\end{align}
\end{subequations}

\textbf{Case d)} 
\begin{subequations}\label{case d 1}  
	\begin{align}
	&\alpha(k) = \frac{A_N }{\gamma_2} \alpha(k-1)  \\
	&\xi_c(k) =  \frac{A_N }{\gamma_2}  \xi_c(k-1)+\frac{F_NC_N}{\gamma_2}\xi_o (k-1)\\
	&\xi_o(k) = \frac{A_N-F_NC_N}{\gamma_2} \xi_o (k-1).
	\end{align}
\end{subequations}

\subsection{Quantized consensus}

Since $(A,C)$ is observable by assumption, we can choose an observer gain $F$ such that $\rho(A-FC) \le \rho(J)$. 
Now we present a technical lemma concerning the upper bound of $\| \alpha(s_r) \|$. 

\begin{lemma}\label{Lemma3}
In view of the scaling parameter (\ref{eq h}), select
\begin{align}\label{34}
\rho(J)< \gamma _ 1<1,\quad \gamma _2 > \rho(A)
\end{align}
and let $\gamma_0:=\max\{\rho(J)/\gamma_1,\rho(A)/\gamma_2\}$. 
Suppose that the quantizer (\ref{quantizer}) does not saturate at successful transmission instants such that $\|\xi_c (s_p)\|_\infty \le \sigma / \gamma_1$ for $p=0,\cdots, r$.
If $\Delta$ and $\tau_D$ regulating DoS frequency satisfy
\begin{align}\label{bound Lemma 3}
\frac{\Delta}{\tau_D} < -\frac{\ln \gamma_0}{\ln C_A C_J}
\end{align}
where $C_A, C_J \in \mathbb R _{\ge 1}$
satisfy $  \|    (  A/\gamma_2 )^k    \| \le C_A  (  \rho(A)/\gamma_2   )^ k$ and 
$
\|   \left(   J/\gamma_1  \right)^k    \| \le C_{J}  \left(  \rho(J)/\gamma_1 \right)^ k$ ($k \in \mathbb Z _{\ge 0}$), respectively, 
then 
\begin{align}
\| \alpha(s_r)    \|  
\le C_3 \sqrt{Nn},
\end{align}
in which
 \begin{align}
C_3 : =\max \left\{2C_1 \frac{C_{x_0}}{\theta_0} , \frac{C_1 C_A \|L\|}{1-\gamma_3 }  \left(\frac{\sigma}{\gamma_1 ^2} + \frac{C_2 C_{x_0}}{\gamma_1 \theta_0}\right) \right\}.
 \end{align}
with $C_1$ and $\gamma_3$ in (\ref{47}), and $C_2$ below (\ref{51}).
\end{lemma}

\emph{Proof.} 
We start the analysis from a successful transmission instant $k-1 = s_{r-1}$ ($k-1 \notin H_q$). If the next instant $k\in H_q$ (corrupted by DoS), one can see that this scenario ($k-1 \notin H_q$ and $k\in H_q$) corresponds to Case c), and one can obtain $\alpha(k)$ by (\ref{case c 1}a).
If $k+1 \in H_q$ as well, this scenario ($k\in H_q$ and $k+1 \in H_q$) corresponds to Case d). Then according to (\ref{case d 1}a), one can obtain $\alpha(k+1)=\frac{A_N}{\gamma_2}\alpha(k)$. 
By iteration, it is straightforward that if all the transmissions at $k+1, \cdots, k+m$ fail, then 
\begin{align}\label{12}\setlength{\arraycolsep}{1pt} 
&\alpha(k+m) 
= \left(\frac{A_N}{\gamma_2}\right)^m \alpha(k)\nonumber\\
&\!\!\!\!\!\!=\left(\!\!\frac{A_N}{\gamma_2}\!\!\right)^m \!\!\!  \left(\!\!  \frac{G}{\gamma_2}\alpha(k-1)       \!  +\! \frac{L}{\gamma_2}\left(\xi_o(k-1)+\xi_c(k-1)\right) \!\! \right)
\end{align}
where the last equality is obtained by substituting $\alpha(k)$ in (\ref{case c 1}a).
If $k+m+1 \notin H_q$ is an instant of successful transmissions, namely $k+m+1=s_r$, then by (\ref{case b 1}a) in Case b) and (\ref{12}), one has that 
$
\alpha(k+m+1) 
 =\frac{A_N}{\gamma_1} \alpha(k+m)=\frac{A_N}{\gamma_1}  \left(\frac{A_N}{\gamma_2}\right)^m \left(  \frac{G}{\gamma_2}\alpha(k-1)         +\frac{L}{\gamma_2}\left(\xi_o(k-1)-\xi_c(k-1)\right)  \right)
$. 
By switching the $\gamma_1$ in $A_N / \gamma_1$ with the $\gamma_2$ in $G /\gamma_2$ and $L/\gamma_2$, and recalling that $k-1 = s_{r-1}$ and $k+m+1=s_r$, 
one has
\begin{align}\label{alpha sr}  
\alpha(s_{r})
&= \left(\frac{A_N}{\gamma_2}\right)^{m_{r-1}}    \frac{G}{\gamma_1}\alpha(s_{r-1})      \nonumber\\   & \quad +\left(\! \frac{A_N}{\gamma_2}\!\!\right)^{m_{r-1}} \!\!\frac{L}{\gamma_1}\left(\xi_o(s_{r-1})+\xi_c(s_{r-1})\right)  
\end{align}  
where $ m_{r-1}\in \mathbb Z _{\ge 0} $ denotes the number of consecutive unsuccessful transmissions between $s_{r-1}$ and $s_r$.
If there is no DoS attack between $s_{r-1}$ and $s_r$, i.e., $m_{r-1}=0$, then (\ref{alpha sr}) is recovered to (\ref{case a 1}a).

There exists a unitary matrix $U$ given by
\begin{align}\label{Phi}
U = [\mathbf{1}/ \sqrt{N} \,\, \phi_2 \,\, \cdots \,\, \phi_N  ] \in \mathbb{R}^{N \times N}
\end{align}
where $\phi_i \in \mathbb{R}^N$ with $i=2, 3, \cdots, N $ satisfies $\phi^T _i L_\mathcal{G} = \lambda_i \phi_i ^ T $ and
$
U^T L_{\mathcal{G}} U = \text{diag}(0, \lambda_2, \cdots, \lambda_N)
$.
With such $U$, we let
\begin{align}
&\overline \alpha(k) :=  (U \otimes I_n)^T  \alpha(k) = \left[ \overline \alpha_1 ^T (k)\,\, \overline \alpha_2 ^T (k)\right] ^T \in \mathbb R^{nN} \nonumber\\
&\bar \xi_c(k) := (U \otimes I_n)^T \frac{L}{\gamma_1} \xi_c(k)= \left[  \bar \xi_{c1} ^T (k)\,\, \bar \xi_{c2} ^T (k)  \right]^T \in \mathbb R^{nN}  \nonumber\\
& \bar \xi_o(k) := (U \otimes I_n)^T \frac{L}{\gamma_1} \xi_o(k)= \left[  \bar \xi_{o1} ^T (k) \,\, \bar \xi_{o2} ^T (k)  \right]^T  \in \mathbb R^{nN}  \nonumber
\end{align}
in which $\bar \alpha_1(k)$, $\bar \xi_{c1} (k)$ and $\bar \xi_{o1} (k) \in \mathbb R^{n}$ denote vectors composed by the first $n$ elements in $\bar \alpha(k)$, $\bar \xi_c (k)$ and $\bar \xi_o(k)$, respectively. 
Then, by (\ref{alpha sr}), one can obtain  
\begin{align}\label{bar alpha 1} 
\bar \alpha(s_{r})
&= 
(U   \otimes I _ n)^T \left(\frac{A_N}{\gamma_2}\right)^{m_{r-1}}  \frac{G}{\gamma_1}  (U \otimes I _n)  \bar \alpha(s_{r-1}) \nonumber\\
&\quad+ 
(U \otimes I _ n )  ^T \left(\frac{A_N}{\gamma_2}\right)^{m_{r-1}}  \frac{L}{\gamma_1}( \xi_o(s_{r-1})+ \xi_c(s_{r-1})) \nonumber\\
&=\left(\frac{A_N}{\gamma_2}\right)^{m_{r-1}} (U \otimes I_n)  ^T  \frac{G}{\gamma_1}  (U \otimes I_n ) \bar \alpha(s_{r-1}) \nonumber\\
&\quad+ 
\left(\frac{A_N}{\gamma_2}\right)^{m_{r-1}}  ( \bar \xi_o(s_{r-1})+ \bar  \xi_c(s_{r-1}))
\end{align} 
in which the last equality is obtained by the fact that $(U \otimes I_n) ^T$ and $I_N \otimes A ^ {m_r - 1 }= A _N ^{m_r -1}$ are commuting matrices:
$
(U \otimes I_n) ^T (I_N \otimes A ^{m_r -1})  
= U^T \otimes A ^{m_r -1} = (I_N \otimes A ^{m_r - 1})(U \otimes I_n)^T.
$
One can verify that $\overline \alpha_1(k) =  0$ for all $k$. 
It is also easy to verify that
$
(U \otimes I_n)^T G(U \otimes I_n)
=\, \text{diag}(A, A-\lambda_2 BK, \cdots, A- \lambda_N BK)
$. Note that $A_N$ and $G$ are block diagonal matrices, thus, from (\ref{bar alpha 1}), one can obtain the dynamics of $\bar \alpha_2(s_{r}) $ as
\begin{align} \label{bar alpha 2} 
\bar \alpha_2(s_{r}) 
&=\left(
\frac{A_{N-1}}{\gamma_2}\right
)^{m_{r-1}} 
   \frac{J}{\gamma_1}  \bar \alpha_2(s_{r-1}) \nonumber\\
&\quad+ 
\left(\frac{A_{N-1}}{\gamma_2}\right)^{m_{r-1}}   ( \bar  \xi_{o2}(s_{r-1}) +  \bar  \xi_{c2}(s_{r-1})) 
\end{align}
in which $A_{N-1}:=I_{N-1}\otimes A$. One can verify that (\ref{bar alpha 2}) also holds for $r=0$. 
By the iteration of (\ref{bar alpha 2}), one can obtain
\begin{align}\label{49}
\bar \alpha_2(s_r) &= \prod_{k=-1}^{r-1} \left(\left(\frac{A_{N-1}}{\gamma_2}\right)^{m_k} \frac{J}{\gamma_1}\right) \bar \alpha_2(s_{-1})  \nonumber\\
& \,\,\,+ \sum_{k=-1}^{r-1} \prod_{p=k}^{r-2} \left(\left(\frac{A}{\gamma_2}\right)^{m_{p+1}} \frac{J}{\gamma_1}\right)  \nonumber\\
&\quad\quad  \times  \left(\frac{A_{N-1}}{\gamma_2}\right)^{m_k}  ( \bar  \xi_{o2}(s_{k}) +  \bar  \xi_{c2}(s_{k})) 
\end{align}
where we let $\prod_{p=k}^{r-2}(\cdot)= I_{n(N-1)}$ when $p=r-1$ in the second line.

Let $T_S(s_r, s_p)$ and $T_U(s_r, s_p)$ ($s_p>s_r$) denote the numbers 
of successful and unsuccessful transmissions within
the interval $[s_r, s_p)$, respectively. 
Note that $C_J$ and $C_A$ as defined after (\ref{bound Lemma 3}) always exist.
Then, in view of the first line in (\ref{49}), one can obtain
\begin{align}\label{47}
&\left\| \prod_{k=-1}^{r-1} \left(\left(\frac{A_{N-1}}{\gamma_2}\right)^{m_k} \frac{J}{\gamma_1}\right) \right\| 
\le (C_A C_{J})^{\eta + \frac{s_r-s_{-1}}{\tau_D}}  \nonumber\\
&\quad \times\left(\frac{\rho(A)}{\gamma_2}\right)^{  T_U(s_{-1},s_r)}\left(\frac{\rho(J)}{\gamma_1}\right)^{  T_S(s_{-1},s_r)} \nonumber\\
&= \underbrace{(C_A C_{J})^\eta}_{C_1} (\underbrace{(C_AC_{J})^{\frac{\Delta}{\tau_D}}\gamma_{0}}_{\gamma_{3}})^{\frac{s_r-s_{-1}}{\Delta}}
\end{align}
in which $\gamma_{0}=\max \{\rho(A)/\gamma_2, \rho(J)/\gamma_1\}<1$ by the selections of $\gamma_1$ and $\gamma_2$, and $\gamma_3<1$ by (\ref{bound Lemma 3}).
Hence, by (\ref{49}), we have
\begin{align}\label{B9}
\|\bar \alpha_2(s_r)\| 
&\le C_1 \gamma_{3} ^{  \frac{s_r- s_{-1}}{\Delta}}  \| \bar \alpha_2(s_{-1}) \| \nonumber\\
&+ C_1 \sum_{p=-1}^{r-1} \gamma_{3} ^{\frac{s_{r-1} - s_{p}}{\Delta}} C_{A}\| \bar  \xi_{o2}(s_{p}) +   \bar  \xi_{c2}(s_{p}) \|
\end{align}
in which
$
\|\overline \alpha_2 (s_{-1}) \| \le   2\sqrt{Nn}  C_{x_0}/\theta_0
$
by (75) in \cite{feng2020arxiv}.
Note that
$
\|\xi_c(s_{-1})\|_\infty 
= 
\|\xi_c(0)\|_\infty 
\le \left\|    (\tilde  x(0) - \hat x(0))/\theta_0   \right\|_\infty 
= \left\|    0/\theta_0    \right\|_\infty =0.
$
By assumption, we have $\|\xi_c (s_p)\|_\infty \le \sigma/\gamma_1 $ for $p=0, 1, \cdots, r$. Incorporating $\|\xi_c(s_{-1})\|_\infty$, overall for $p=-1, 0, \cdots, r$, one has 
$
\|\xi_c (s_p)\|_\infty \le \sigma/\gamma_1
$
and hence
\begin{align}\label{50}
&\|\bar \xi_{c2} (s_p)\| \le  \|\bar \xi_c(s_p) \| \le \|L\| \|\xi_c(s_p)\| / \gamma_1  \nonumber\\
   & \quad 
  \le \|L\| \sqrt{Nn} \|\xi_c(s_p)\|_\infty / \gamma_1 \le \|L\| \sqrt{Nn} \sigma /\gamma_1^2.
\end{align}

By the dynamics of $\xi_o(k)$ in Cases a)--d), we have
\begin{align}\label{51}
\|\xi_o(k)\| &\le C_2 (\rho(A-FC)/\gamma_1)^k \|\xi_o(0)\| \nonumber\\
&< C_2 \|\xi_o(0)\| \le  \sqrt{Nn} C_2 C_{x_0} /\theta_0, 
\end{align}
in which $C_2 \ge 1$, $\rho(A-FC)/\gamma_1 \le \rho(J)/\gamma_1<1$ by the selection of $F$ and $\|\xi_o(0)\| = \|\frac{x(0)-\hat x(0)}{\theta_0}\|=\|\frac{x(0)}{\theta_0}\|\le \sqrt{Nn}C_{x_0}/\theta_0$. By (\ref{51}), one has 
\begin{align}\label{52}
\| \bar \xi_{o2}(s_p)\| & \le\| \bar \xi_{o}(s_p)\| \le  \|L\| \|\xi_o(s_p)\| / \gamma_1  \nonumber\\
&\le \|L\| \sqrt{Nn}  C_2 C_{x_0} /(\theta_0 \gamma_1).
\end{align}

Note that in (\ref{B9}), $(s_r - s_{-1})/\Delta \ge r$ and $(s_r - s_{k+1})/ \Delta  \ge r-k$ with $k=-1, \cdots, r$. 
Substituting (\ref{50}) and (\ref{52}) into (\ref{B9}), we have
\begin{align}\label{upper bound}
&\|\overline \alpha_2 (s_{r}) \|  \le C_1  \gamma_3    ^r 2\sqrt{Nn} \frac{C_{x_0}}{\theta_0} \nonumber\\
& \quad \quad\quad \quad\,\, + C_1 C_A \frac{1-  \gamma_3  ^r}{1-\gamma_3 } \|L\| \sqrt{Nn}\left(\frac{\sigma}{\gamma_1 ^2} + \frac{C_2 C_{x_0}}{\gamma_1 \theta_0}\right)   \nonumber\\
&\! \le \underbrace{\max \left\{\! 2C_1 \frac{C_{x_0}}{\theta_0} , \frac{C_1 C_A \|L\|}{1-\gamma_3 }  \left(\! \frac{\sigma}{\gamma_1 ^2} + \frac{C_2 C_{x_0}}{\gamma_1 \theta_0}\right) \! \right\}  } _{C_3} \sqrt{Nn}.
\end{align}
Then, one has 
$
\|\alpha(s_r)\| \le \|( U   ^T) ^ {-1}\| \|\overline \alpha(s_r)\|  
=\|\overline \alpha_2(s_r) \|  \le C_3 \sqrt{Nn} $. Trivially, (\ref{upper bound}) also holds for $r=-1$.
\qedp

\begin{remark}\label{remark 2}
	The upper bound of $\alpha(s_r)$ is influenced by the frequency of DoS attacks shown by (\ref{bound Lemma 3}), and therefore influences $[\alpha ^ T (s_r)\,\,\xi_c ^ T (s_r)\,\,\xi_o ^ T (s_r) ]^T$, which determines the necessary data rate (see Theorem 1 later). 
	DoS frequency influences $\alpha(s_r)$ due to the nature of switching in the system, e.g., $\alpha(s_r)$ can diverge under fast switches among Cases a)-d) due to fast off/on of DoS attacks. 
	As will be shown later, in case of scalar multi-agent systems, DoS-frequency constraint in (\ref{bound Lemma 3}) does not influence the problem any more. \qedp
\end{remark}

\begin{remark}
We emphasize that $\gamma_2$ in Lemma \ref{Lemma3} is lower bounded by $\rho(A)$, and it is tighter than that in \cite{feng2020arxiv}. We explain how the new controller can attain this. One of the major functions of the new controller is to obtain Case d), in which $\alpha(k)$ and $\xi_c(k)$ in (\ref{case d 1}) are decoupled under DoS and regulated by $A_N$. If one adopts the controller in \cite{feng2020arxiv}, $\alpha(k)$ and $\xi_c(k)$ are coupled under DoS attacks. This implies that one can not simply multiply $\alpha(k)$ by $(A_N /\gamma_2)^m$ in (\ref{12}). Instead, one needs to simultaneously consider $\alpha(k+m)$ and $\xi_c(k+m)$ due to the couplings, and obtains a form similar to (\ref{bar alpha 1}) but $(A_N/\gamma_2) ^ {m_r -1}$ should be replaced by a matrix derived from the combinations of $G, L$, $H$ and $\gamma_2$. Then, to obtain $\bar \alpha_2(s_r)$ from $\bar \alpha(s_r)$, one needs to remove some lines and rows of the matrix \cite{feng2020arxiv}, and hence $\gamma_2$ loses its connections to $A$. Eventually, in order to obtain a $\gamma_2$ realizing zooming-out, one needs to take the worst-case analysis (i.e., consider all the possible scenarios of consecutive packet losses and compute the corresponding $\gamma_2$, then select the largest $\gamma_2$), and this is the major reason of conservative and inexplicit $\gamma_2$ in \cite{feng2020arxiv}. This is also computationally intense.    \qedp
\end{remark}

\begin{remark}
	Compared with the results in \cite{feng2020arxiv}, the parameters $\gamma_1$ and $\gamma_2$ in this paper can be directly selected when $J$ and $A$ are known, respectively. In contrast, the choices of $\gamma_1$ and $\gamma_2$ in \cite{feng2020arxiv} need complex calculations including matrix multiplication and taking their norms for as many rounds as the number of maximum consecutive packet losses (see Lemma 3, \cite{feng2020arxiv}). 
	That is to say, one needs to know the maximum number of consecutive packet drops in order to compute $\gamma_2$. 
	By contrast, to obtain $\gamma_2$ in this paper, such information is not needed and one only needs to compute $\rho(A)$ and then selects a tight value provided that (\ref{bound Lemma 3}) holds. To attain the above merits of the new controller over those in \cite{feng2020arxiv}, one needs to deal with a stabilization problem of a switched system of four modes (i.e, Cases a)-d)) in the technical analysis, instead of two modes in \cite{feng2020arxiv}. As will be shown later in the proof of Theorem \ref{Theorem 1}, the four-mode switched system also complicates the calculation of data rate.   \qedp
\end{remark}

In general, it is difficult to tighten the zooming-out factor beyond the eigenvalues of open-loop systems. This aspect can be more clearly discussed for centralized systems. For example, in the paper \cite{you2010minimum} considering a discrete-time system, the quantization needs to zoom out with the factor of $|\lambda^u_i(A)|$ (the $i$-th unstable eigenvalue of $A$) for each failed transmission in order to ``catch" the diverging estimation error. Similarly, for a continuous-time system, the zooming-out factor is continuous $e^{\text{Re} \lambda^u_i(A)t}$ (real part of $\lambda^u_i(A)$: Re$\lambda^u_i(A)>0$) during open-loop intervals \cite{feng2020tac}. 
For multi-agent systems, the lack of global state is another challenge. The results mentioned for centralized systems significantly depend on designing a fine predictor having access to the entire state. However, such a predictor is not applicable to multi-agent systems because of the distributed system structure where the agents are constrained to have only local information of their direct neighbors. Therefore, in \cite{feng2020arxiv}, the authors attempt to predict the state by an open-loop estimator and feed the estimations to the feedback controller. However, due to matrix manipulations, the zooming-out factor eventually loses its connections to the eigenvalues of $A$.

In the following, we present the main result of the paper.

\begin{theorem}\label{Theorem 1}
	Consider the multi-agent system (\ref{system}) with control inputs (\ref{local observer}) to (\ref{eq h}), where $\gamma_1$ and $\gamma_2$ are selected as in Lemma \ref{Lemma3}. Suppose that the DoS attacks characterized in Assumptions 1 and 2 satisfy (\ref{bound Lemma 3}). 
	 If $2R+1$ in the quantizer (\ref{quantizer}) satisfies
	\begin{align}
	(2R+ 1 ) \sigma \ge  \zeta C_5 \sqrt{Nn} \label{Theorem quantizer}  
	\end{align}
where
$
C_5:=((C_3\|L\|+ \|H\| \sigma/\gamma_1 + \|P\|C_2C_{x_0}/\theta_0)^2  
+ \|A-FC\|^2 C_2 ^2 C_{x_0}^2/\theta_0 ^2)^{\frac{1}{2}}
$
and $\zeta: = \max\{1, C_4\|A\,\,\, FC\|_\infty /\gamma_2\}$, then quantizer (\ref{quantizer}) is not overflowed. 
Moreover, if DoS attacks satisfy
\begin{align}\label{DoS bound}
\frac{1}{T} + \frac{\Delta}{\tau_D} < \frac{-\ln \gamma_1}{\ln \gamma_2 -  \ln \gamma_1}
\end{align}
then consensus of $x_i(k)$ is achieved as in (\ref{control objective}). The parameter $C_4 \ge 1$ in $\zeta$ is chosen such that $\|(S/\gamma_2) ^{m-1}\| \le C_4 \rho(S/\gamma_2)^{m-1}\le C_4$ ($m \in \mathbb Z _{\ge 1}$), where
\begin{align}
S:=
\left[
	\begin{array}{cc}
A_N & F_N C_N\\
0  & A_N -F_NC_N
\end{array}
\right].
\end{align}
\end{theorem}


\emph{Proof.}  
The unsaturation of quantizer is proved by induction: if the quantizer satisfying (\ref{Theorem quantizer}) is not overflowed such that $\|\xi_c (s_p)\|_\infty \le \sigma / \gamma_1 $ for $p=0, \cdots, r$ and recall that $\|\xi_c (s_{-1})\|_\infty=0 \le \sigma / \gamma_1 $, then the quantizer will not saturate at the transmission attempts within the interval $(s_r, s_{r+1}]$ and hence $\|\xi _c (s_{r+1})\|_\infty \le \sigma / \gamma_1 $. 

\emph{a)} In the proof, $s_r$ represents an instant of successful transmission ($r \in \mathbb Z _{\ge0}$) or the initial time $s_{-1}$. At $s_r + 1$, the quantized information of $\hat x(s_r + 1)$ attempts to transmit through the network, i.e., 
$
Q_R\left( (\hat  x(s_r + 1) - A_N \tilde  x(s_r)  )       /\theta (s_r) \right). 
$
In order to prevent quantizer overflow, $\| (\hat x(s_r + 1) - A_N \tilde x(s_r)  )/\theta (s_r) \|_\infty $ must be upper bounded by the maximum quantization range, i.e., $(2R+1)\sigma$. By (\ref{HK}), one has
\begin{align}
&\frac{\hat x(s_r + 1) - A_N \tilde x(s_r)}{\theta (s_r)}= \frac{He_c(s_r) - L \delta(s_r) + P e_o(s_r)  }{\theta(s_r)} \nonumber\\
&= H \xi_c (s_r) - L \alpha(s_r) + P \xi_o(s_r). 
\end{align}  
One can verify that the quantizer at $s_r + 1$ is not saturated since 
$
\left\| [-L\,\, H\,\, P]   
[
\alpha^T(s_r)\,\, 
\xi_c ^T (s_r)\,\, \xi_o^T(s_r)
]^T
\right\|_\infty 
\le (2R + 1) \sigma
$ in (\ref{Theorem quantizer}), where
$\|\alpha(s_r)\| \le C_3 \sqrt{Nn}$ by Lemma \ref{Lemma3}, $\|\xi_c(s_r)\|\le \sqrt{Nn} \sigma /\gamma_1$ and $\|\xi_o(s_r)\| \le \sqrt{Nn} C_2 C_{x_0} /\theta_0$ in (\ref{51}).

\emph{b)} At $s_r + 2$, quantized signals of $\hat x(s_r + 2)$ attempt to be transmitted to the decoders, and one needs to compute 
$\|
(\hat x(s_r + 2) - A_N \tilde  x(s_r+1))/\theta (s_r+1) \|_\infty . 
$
However, one cannot compute it directly as in \emph{a)} because $\hat x(s_r + 2)$ has two cases: the transmission attempts at $s_r+1$ in \emph{a)} are successful or corrupted by DoS. 
\emph{b-1)} If $s_r + 1 \notin H_q$, then one can apply the similar analysis in \emph{a)} and concludes that $Q_R(\cdot)$ does not saturate at $s_r + 2  $.
\emph{b-2)} If the transmission attempts at $s_r + 1$ in \emph{a)} are not received by the decoders, then at $s_r + 2 $, by (\ref{HK}) with $u(s_r + 1)=0$, one can compute that 
\begin{align}\label{60}
& (\hat x(s_r + 2) - A_N \tilde x(s_r+1))/\theta (s_r+1) \nonumber\\
& = A_N \xi_c (s_r + 1) + F_NC_N \xi_o(s_r + 1) \nonumber\\
& =   \left[A_N \,\, F_N C_N \right] 
\left[
\begin{array}{ll}
\xi_c ^T(s_r+1) &
\xi_ o ^T (s_r+1 )
\end{array}
\right]^T
\end{align} 
in which by (\ref{case c 1}) one has 
\begin{align}\label{58}
&\left[
\begin{array}{ll}
\xi_c ^T(s_r+1) &
\xi_ o ^T (s_r+1 )
\end{array}
\right]^T \nonumber\\
& = 
\frac{1}{\gamma_2}\left[ \begin{array}{ccc}
-L &H& P \\
0 & 0& A_N -F_NC_N 
\end{array}
\right] 
\left[
\begin{array}{l}
\alpha(s_r)\\
\xi_c (s_r ) \\
\xi_ o  (s_r ) 
\end{array}
\right].
\end{align}
By (\ref{60}), (\ref{58}) and the upper bounds of $\|\alpha(s_r)\|, \|\xi_c(s_r)\|$ and $\|\xi_o(s_r)\|$ in \emph{a)}, one can verify that $\| (\hat x(s_r + 2) - A_N \tilde x(s_r+1))/\theta (s_r+1)\|_\infty \le (2R+1)\sigma$ in (\ref{Theorem quantizer}).

\emph{c)} By \emph{b-1)}, if the previous step is not under DoS, one can always follow \emph{a)} to verify quantizer unsaturation. Hence, we omit this case and analyze consecutive packet losses $\{s_r + 1 \cdots s_r+m\}\in H_q$ until $s_r+m+1=s_{r+1}$. At $s_r+m+1$, one should focus $ (\hat x(s_r+m+1) - A_N \tilde x(s_r+m))/\theta (s_r+m)= [A_N \,\, F_N C_N] [\xi_c ^T(s_r+m) \,\,
\xi_ o ^T (s_r+m )]^T$, in which by (\ref{case d 1}) one can obtain 
\begin{eqnarray}\setlength{\arraycolsep}{1pt} \label{eq 60}
	\left[\!
	\begin{array}{ll}
		\xi_c (s_r+m) \\
		\xi_ o  (s_r+m )
	\end{array}
	\!\right] 
	\!= \!
	\frac{S^{m-1} }{\gamma_2 ^{m-1}}
	\left[\!
	\begin{array}{ll}
		\xi_c (s_r+1) \\
		\xi_ o  (s_r+1 )
	\end{array}
	\!\right]
\end{eqnarray}
with $[	\xi_c ^T (s_r+1) \,\, \xi_ o ^T (s_r+1 )]^T$ in (\ref{58}). 
Substituting (\ref{58}) and (\ref{eq 60}) into $[A_N \,\, F_N C_N] [\xi_c ^T(s_r+m) \,\,
\xi_ o ^T (s_r+m )]^T$, one can verify that quantizer is not saturated at $s_r+m+1=s_{r+1}$ by 
\begin{align}\label{61}
& \|[A_N \, F_NC_N ] 
[
\xi_c ^T(s_r+m)\,\,
\xi_ o ^T (s_r+m )
]^T \| _\infty   \nonumber\\
& \le \frac{C_4 \|[A_N \,\, F_N C_N ]\| _\infty }{\gamma_2} 
\left\| 
\left[\!\!\!\!
\begin{array}{c}
-L \alpha(s_r) + H \xi_c(s_r) + P \xi_o(s_r) \\
(A_N -F_N C_N)\xi_o(s_r )
\end{array}\!\!\!\!
\right]
 \right\| \nonumber\\
& \le   \frac{C_4 \|[A_N \,\, F_N C_N ]\| _\infty }{\gamma_2} ((\|L\| \|\alpha(s_r)\| + \|H\| \|\xi_c(s_r)\| \nonumber\\
&  \quad  + \|P\| \|\xi_o(s_r)\|    )^2 + \|A_C-F_NC_N\|^2 \|\xi_o(s_r)\|^2 )^{1/2} \nonumber\\
&  \le \zeta C_5 \sqrt{Nn} \le (2R +1) \sigma
\end{align}
where $C_4\ge 1$ defined below (\ref{DoS bound}) exists since $\rho(S/\gamma_2)<1 $.

By the analysis in \emph{a)}--\emph{c)}, one can conclude that the quantizer does not saturate during $ [s_{-1}+1, s_0]\bigcup\, (s_{r}, s_{r+1}]$ with $r \in \mathbb Z _{\ge0}$. This implies that the quantizer does not saturate at all $k$.

 \emph{d)}
 Now we show state consensus. We first need to prove that $\|\alpha(k)\|$ is finite for all $k$. For this, it is sufficient to show $\|\alpha(k)\|$ is bounded during $(s_r, s_{r+1}]$. We have proved that $\|\alpha(s_r)\|$ and $\|\alpha(s_{r+1})\|$ are upper bounded by Lemma \ref{Lemma3}. Then, we only need to show $\|\alpha(\cdot)\|$ is bounded for $s_r<s_r+1, \cdots, s_r+m<s_r+m+1=s_{r+1}$. If $s_r +1$ is a failed transmission instant, by (\ref{case c 1}a), one can infer that $\|\alpha(s_r + 1)\|$ is upper bounded. One can also infer that $\|\alpha (s_r + m )\|$ is also upper bounded in view of $\alpha(s_r + m ) = (A_N/\gamma_2)^{m-1} \alpha(s_r +1)$ by (\ref{case d 1}a) with $m\in \mathbb N_{\ge 2}$ and $\rho(A_N/\gamma_2)<1$. By the analysis above, we conclude that all the $\|\alpha(k)\|$ during $(s_r, s_{r+1}]$ is upper bounded and hence $\alpha(k)$ is finite for all $k$. 
 Recall the definitions of $T_S(1, k)$ and $T_U(1, k)$ before Lemma \ref{Lemma T}. In view of $\delta(k) = \theta(k) \alpha(k)  = \gamma_1 ^ {T_S(1,k)} \gamma_2 ^{T_U(1,k)}  \theta_0 \alpha(k) $, one has
$
 	\|\delta(k)\| \le  C_3  \gamma^k  \theta_0 \|\alpha(k)\| 
$
 where $C_3 = \left( \gamma_2 /\gamma_1   \right)^{(\kappa+\eta\Delta)/\Delta} $ and 
$
 	\gamma = \gamma_1^{1-\frac{1}{T}-\frac{\Delta}{\tau_D}  } \gamma_2^{ \frac{1}{T}+ \frac{\Delta}{\tau_D}   } < 1
$
 by (\ref{DoS bound}). Thus, we have $\|\delta(k)\|  \to 0$ as $k\to \infty$, which implies state consensus. 
 \qedp

 \color{black}

\begin{remark}
In principle, one can select $\gamma_1$ arbitrarily close to $\rho(J)$ and $\gamma_2$ close to $\rho(A)$ in order to improve the system resilience for reaching consensus in view of (\ref{DoS bound}). Such choices are effective especially under long time but less frequent DoS attacks. However, such choices may also lead to larger $C_A$ and $C_J$, respectively, and a smaller $\gamma_{0}$ in (\ref{bound Lemma 3}). This implies that $\alpha(s_r)$ may diverge under frequent DoS attacks due to fast switches in a switched control system (see Remark \ref{remark 2}). It is clear that a small $\Delta$ can always relax the constraints in (\ref{bound Lemma 3}) and (\ref{DoS bound}), but can increase the communication burden.      \qedp
\end{remark}

\subsection{Scalar multi-agent systems}
If $A\in \mathbb R ^{n \times n}$, one sees that overflow problem of quantizer is subject to dwell time constraint in (\ref{bound Lemma 3}). As mentioned before, the system is not subject to the constraint in case $A \in \mathbb R$, i.e. a scalar multi-agent system. More importantly, in case $A \in  \mathbb R$, we are able to further tighten the zooming-out factor, i.e., smaller than $|A|$ and therefore recover the robustness result of unquantized control, i.e., 
if DoS attacks satisfy
\begin{align}\label{optimal} 
\frac{1}{T} + \frac{\Delta}{\tau_D} < \frac{-\ln \rho(J)}{\ln \rho(A)- \ln \rho(J)} ,
\end{align}
consensus of $x_i(k)$ is achieved and the quantizer is not saturated. We briefly present the proof of unquantized case obtaining (\ref{optimal}) in the Appendix. In the following, we present the result of quantizer unsaturation and consensus for scalar multi-agent systems. 
We assume that $|A|>1$ and $x_i(k)$ is directly known, since one can always obtain $x_i(k)=y_i(k)/C$.  


The controller in (\ref{controller}) to (\ref{eq h}) is still applicable, and $\hat x_j(k)$ in (\ref{Q_i}) should be replaced by $x_j(k)$. Consequently, $e_o(k)=0$ and $\xi_o(k)=0$ for all $k$.

\begin{proposition}\label{Theorem 2}
	Consider the multi-agent system (\ref{system}) with $A\in \mathbb R$ and the control inputs (\ref{controller}) to (\ref{eq h}). Suppose that the DoS attacks characterized in Assumptions 1 and 2 satisfy $1/T + \Delta/\tau_D < 1$.  
	For any $\rho(J)<\gamma_1<1$, the choice of $\gamma_2$ should satisfy 
	\begin{align}\label{gamma 2 DoS}
	\ln \gamma_2 = \ln \gamma_1 \ln A /\ln \rho(J),
	\end{align}
	then the followings hold:
	\begin{itemize}
\item[(1)] 	the quantizer does not saturate if $(2R+1)\sigma \ge \zeta \|L \,\,H\|_\infty C_7 \sqrt{N}$ ($C_7$ and $\zeta$ are positive given in proof); 
\item[(2)] 	Moreover, if (\ref{optimal}) holds, consensus of $x_i(k)$ is achieved. 
	\end{itemize}
\end{proposition}

\emph{Proof.}
Note that in order to prove consensus by showing $\|\delta(k)\|\to 0$, one can follow the analysis in Lemma \ref{Lemma3} and conclude $\|\alpha(s_r)\|=\|\bar \alpha_2(s_{r})\|$ is finite. 
One can obtain the dynamics of $\bar \alpha_2(s_{r}) $ similar to (\ref{bar alpha 2}), in which $\bar \xi_{o2}(s_{r-1})=0$ for all $s_{r-1}$. Then the counterpart of (\ref{47}) is given by 
\begin{align}\label{scalar 60}
& \left\|
(A_{N-1}/\gamma_2 )^{  T_U(s_{-1},s_r)}
(J/\gamma_1)^{  T_S(s_{-1},s_r)}
\right\|
\nonumber\\
& \le \underbrace{\left(\frac{\rho(A) \gamma_1}{\rho(J)\gamma_2}\right)^{\frac{\kappa + \eta \Delta}{\Delta}} }
_{C_6}
\bigg(
\underbrace{\left(
	\frac{\rho(A)\gamma_1}{\rho(J)\gamma_2}
	\right)^{\frac{1}{T}+ \frac{\Delta}{\tau_D}}
	\frac{\rho(J)}{\gamma_1}}
_{\gamma_4}
\bigg)
^{\frac{s_r-s_{r-1}}{\Delta}} \!\!\!\!
\end{align}  
and the counterpart of (\ref{B9}) is given by 
$
\|\bar \alpha_2(s_r)\| 
\le C_6 \gamma_4 ^{  \frac{s_r- s_{-1}}{\Delta}}  \| \bar \alpha_2(s_{-1}) \|
+ C_6 \sum_{k=0}^{r} \gamma_4 ^{\frac{s_r - s_{k-1}}{\Delta}} \|\bar \xi _{c2}(s_{k-1})\|.
$
As $A_{N-1}$ and $J$ are diagonal matrices, the stability of the switched system is free from the dwell time constraint in (\ref{bound Lemma 3}). In order to ensure the boundedness of $\bar \alpha_2(k)$, $\gamma_4$ needs to be smaller than 1, which is implied by 
\begin{align}\label{64}
	\frac{1}{T} + \frac{\Delta}{\tau_D} <
	\frac{- \ln  (\rho(J)/\gamma_1 }{\ln (A/\gamma_2 )-  \ln  (\rho(J)/\gamma_1)}.
\end{align}
The bounds of $\|\bar \alpha_2(s_{-1})\|$ and $\|\bar \xi_{c2} (s_{k-1})\|$ obtained in the proof of Lemma \ref{Lemma3} still hold. Then, one can obtain $\|\alpha(s_r)\|\le \sqrt{N}C_7$ with $C_7:=\max\{2C_6C_{x_0}/\theta_0, C_6\sigma /(\gamma_1^2 (1-\gamma_4))\}$.

At $s_r + 1$, the quantized information of $x(s_r + 1)$ attempts to transmit
$
Q_R\left( ( x(s_r + 1) - A_N \hat x(s_r)  )       /\theta (s_r) \right)$, and hence $\| (x(s_r + 1) - A_N \hat x(s_r)  )/\theta (s_r) \|_\infty $ must be upper bounded by the maximum quantization range. Specifically, one has
$
\!\!(x(s_r + 1) - A_N \hat x(s_r))/\theta (s_r)= (H e_c(s_r) - L \delta(s_r) )/\theta(s_r) \nonumber\\
=  H \xi_c(s_r) - L \alpha(s_r),
$
and the quantizer at $s_r + 1$ is not saturated since 
$
\| [-L \quad H\,]   
\left[
\begin{array}{ll}
\alpha^T(s_r) 
\xi^T _c (s_r)
\end{array}
\right]^T
\|_\infty 
\le (2R + 1) \sigma. 
$

At $s_r + 2$, the quantized signals of $x(s_r + 2)$ needs to be transmitted to the decoders, and one needs to compute 
$\|
(x(s_r + 2) - A_N \hat x(s_r+1))/\theta (s_r+1) \|_\infty. 
$
If the transmission attempts at $s_r + 1$ are successfully received by the decoders, then one can apply the similar analysis in the former paragraph and then concludes that $Q_R(\cdot)$ does not encounter the overflow problem at $s_r + 2 $.
If the transmission attempts at $s_r + 1$ are not received by the decoders, then at $s_r + 2 $, one can obtain
	\begin{align}
	& \frac{x(s_r + 2) - A_N \hat x(s_r+1)}{\theta (s_r+1)} \nonumber\\
	& = \frac{ A_N  x(s_r + 1)    -A_N  \hat x(s_r + 1 )}{\theta (s_r + 1)}  = \frac{  A_N  e_c(s_r + 1)  }{\theta (s_r + 1)} \nonumber\\
	& = A_N \xi_c (s_r + 1)  = \frac{A_N}{\gamma_2} \left[ -L \,\, H \right]  
[
	\alpha^T(s_r) \,\,
	\xi_c ^T (s_r)
]^T
	\end{align} 
	where the last equality is due to 
	$
	\xi _c (s_r + 1) = \frac{H}{\gamma_2} \xi _c (s_r) - \frac{L}{\gamma_2} \alpha(s_r)
	$
	according to (\ref{case c 1}b), in which $\xi_o(k-1)=0$.
By induction, if all the transmission attempts at $s_r + 1, \cdots, s_r + m $ fail, then one can obtain that at $s_r + m +1= s_{r+1} \notin H_q$, the quantizer does not saturate since 
\begin{align}\label{65}
& \| 
(x(s_r + m+1) - A_N \hat x(s_r+ m) )   /   \theta (s_r+m)
\| _ \infty \nonumber\\
& =\left\| 
( A_N /\gamma_2)^{m} [ -L \,\, H] 
[
\alpha^T(s_r) \,\,\,
\xi_c ^T (s_r)
]
^T  \right\|_\infty
\nonumber\\
& \le (2R+1) \sigma
\end{align} 
in which $\|(A_N/\gamma_2)^{m}\| \le (\max\{1, A/\gamma_2 \})^M =:\zeta$. Here, $M\in \mathbb Z_{\ge 0}$ denotes the maximum number of consecutive packet losses, and can be calculated by Lemma 2 in \cite{feng2020arxiv}.

For showing consensus, one needs to prove $\|\delta(k)\|=\theta(k)\|\alpha(k)\|\to 0$ as $k\to \infty$, in which $\|\alpha(k)\|$ can be shown upper bounded by following an analysis similar to that in Theorem \ref{Theorem 1}. 
By (\ref{DoS bound}), it holds that $\theta(k) \to 0$ as $k \to \infty$.
Overall, in order to achieve consensus and quantizer unsaturation simultaneously, DoS attacks need to satisfy 
\begin{eqnarray}\label{67}
\frac{1}{T} + \frac{\Delta}{\tau_D} < \min \left\{\frac{-\ln \gamma_1}{\ln \gamma_2 - \ln \gamma_1},  \frac{- \ln \rho(J)/ \gamma_1 }{\ln A/\gamma_2 - \ln \rho(J)/\gamma_1} \right\}
\end{eqnarray}
where the second term in $\min\{\cdot\}$ is obtained by imposing $\gamma_4<1$ in (\ref{64}).
One can verify that since $\rho(J)<\gamma_1<1$ and $\gamma_2$ satisfy (\ref{gamma 2 DoS}),
it holds 
$
\frac{-\ln \gamma_1}{\ln \gamma_2 - \ln \gamma_1}=   \frac{- \ln \frac{\rho(J)}{ \gamma_1}  }{\ln \frac{A}{\gamma_2} - \ln \frac{\rho(J)}{\gamma_1}} =
\frac{-\ln \rho(J)}{\ln A - \ln \rho(J)}
$
which leads to (\ref{optimal}). 
%
\qedp

\begin{remark}\label{remark 3}
In order to preserve (\ref{optimal}), given a $\gamma_1$, one needs to compute $\gamma_2$ by (\ref{gamma 2 DoS}). Otherwise, (\ref{optimal}) cannot be preserved. One should notice that $\gamma_2$ obtained by (\ref{gamma 2 DoS}) is smaller than $A$ since $\ln \gamma_1/ \ln \rho(J)>1$. As a result, to obtain Proposition \ref{Theorem 2}, we need the information of the maximum consecutive packet drops ($M$) in order to compute $\zeta$ and therefore the necessary data rate by (\ref{65}). However, if one lets $\gamma_2=A$, then the information of $M$ is not needed, while one could not recover (\ref{optimal}), which implies a less robust system. 
One can verify that the right-hand side of (\ref{optimal}) is always larger than that of (\ref{DoS bound}), which implies a better robustness of a scalar multi-agent system than a general linear multi-agent system. 
\qedp
\end{remark}

\begin{figure}[t]
	\begin{center}
		\includegraphics[width=0.4\textwidth]{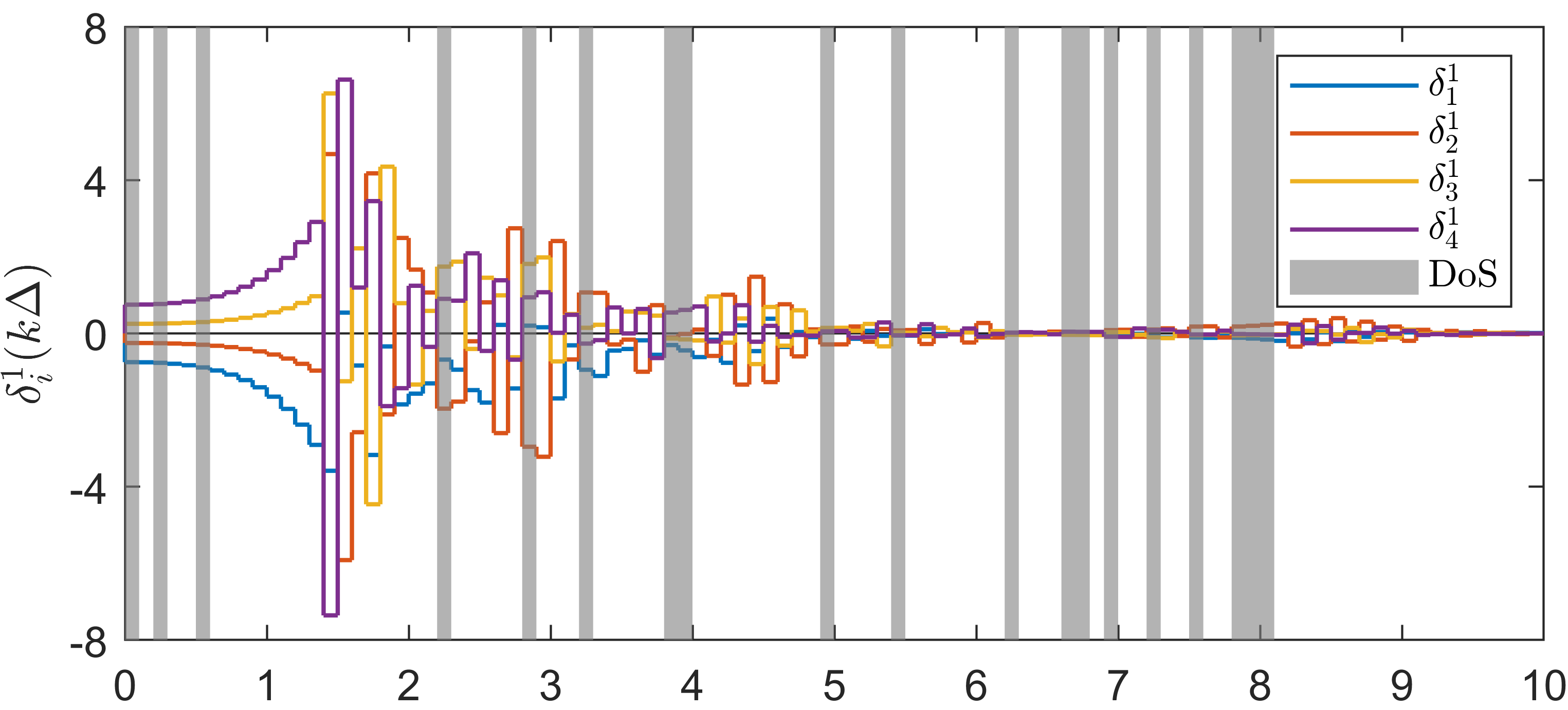}  \\
		\includegraphics[width=0.4\textwidth]{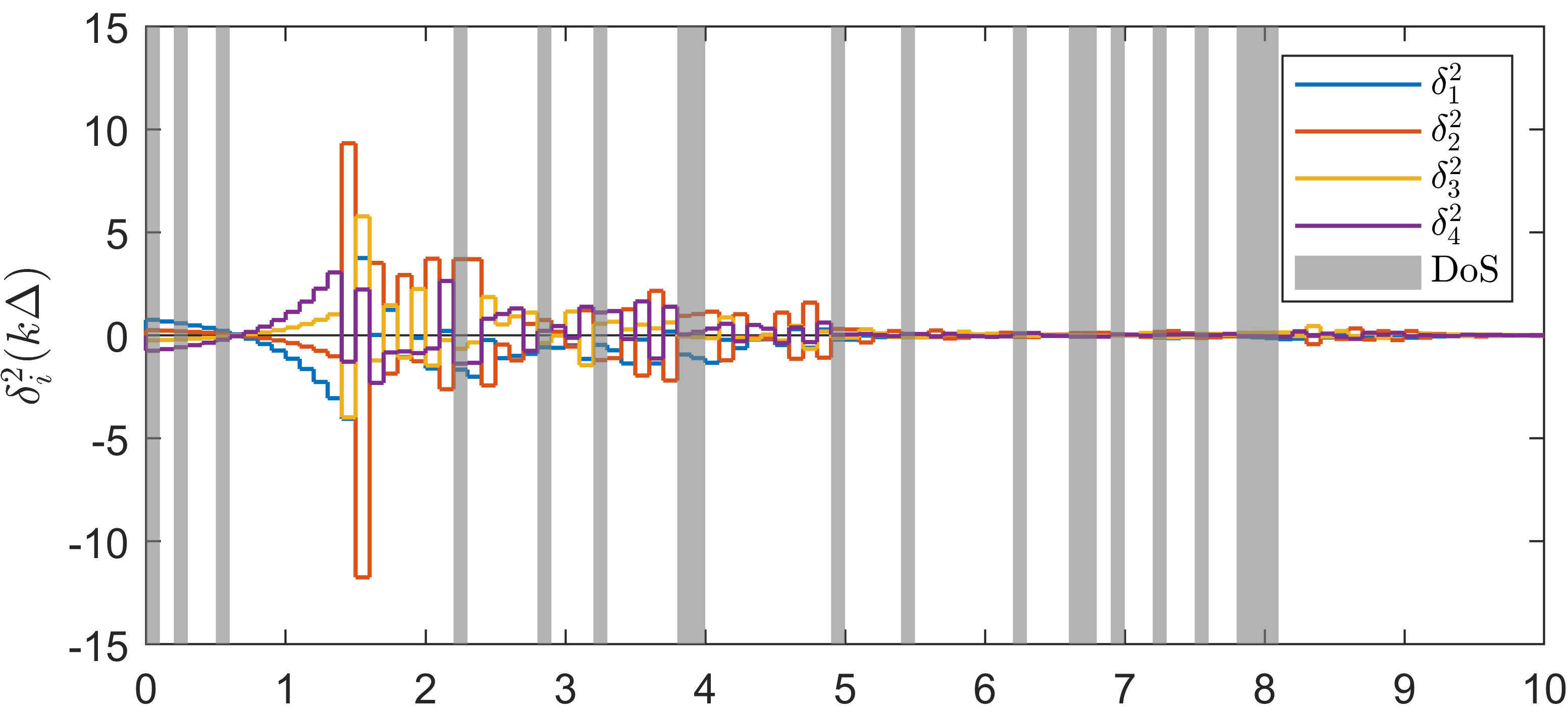}  \\
		\includegraphics[width=0.4\textwidth]{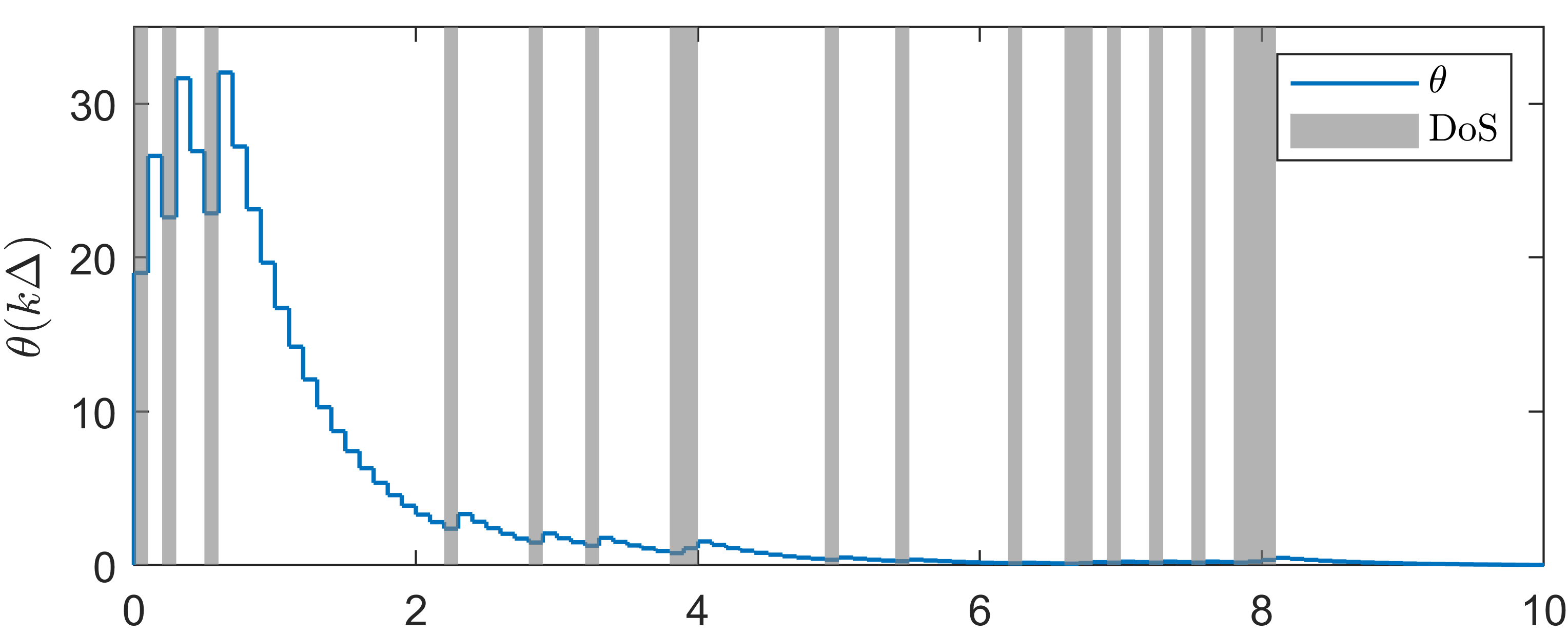}  \\
		\includegraphics[width=0.4\textwidth]{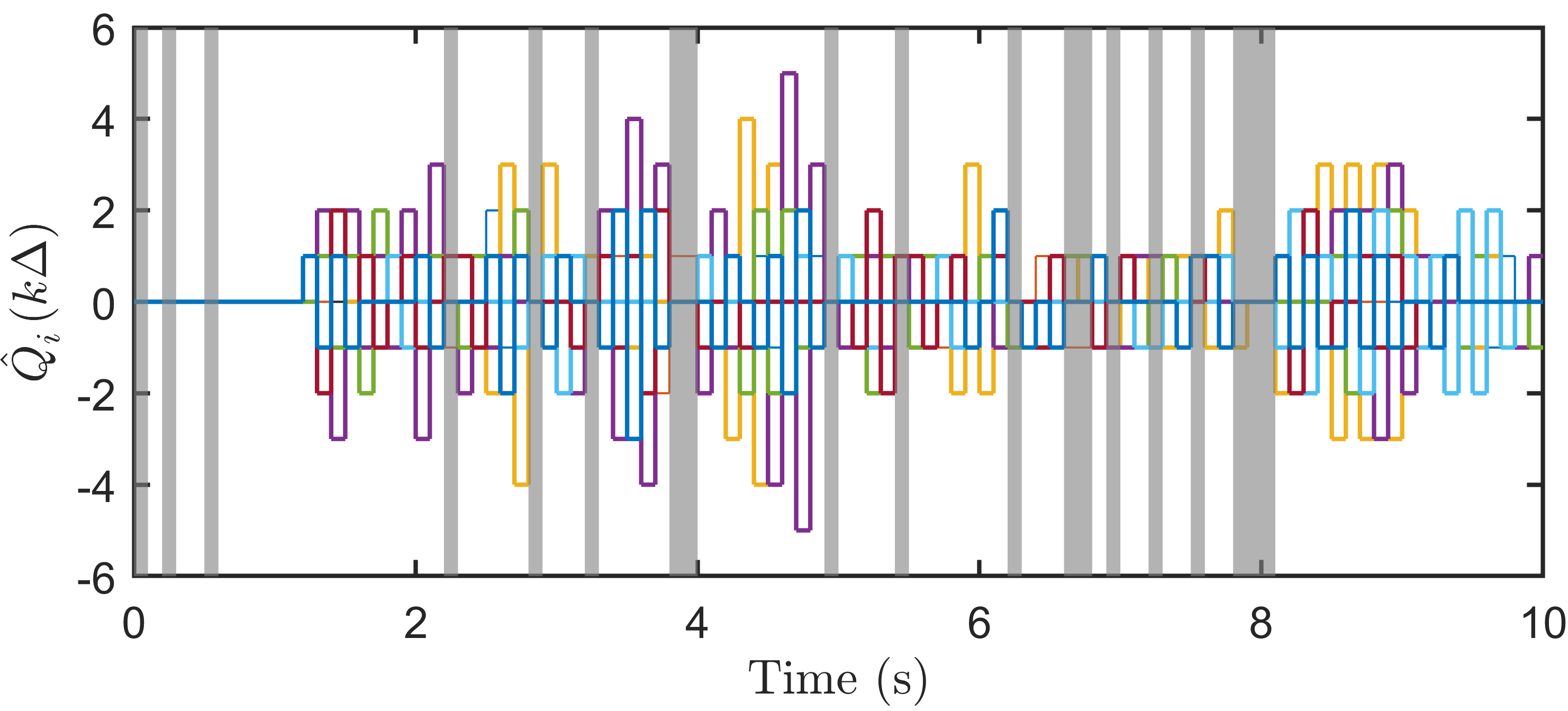}  \\
		\vspace{-2mm}
		\linespread{1}\caption{Time responses of $\delta_i ^1 (k)$ (first plot), $\delta_i ^2 (k)$ (second plot), $\theta(k)$ (third plot) and $\hat Q _i (k)$ (last plot). Note $\delta_i(k)=[\delta_i^1(k) \,\, \delta_i^2(k)]^T$ ($i=1, 2, 3, 4$). }\label{simulation1}
	\end{center}
\vspace{-8mm}
\end{figure}

\section{Numerical example}
In the simulation example, we consider a multi-agent system of $N=4$ with
\begin{align}\setlength{\arraycolsep}{2pt} 
A= 
\left[\begin{array}{cc}
1.1162  &  0.1109 \\
0.2218   & 1.1162
\end{array}
\right],\,
B= 
\left[\begin{array}{cc}
0.1052  &  0.0053 \\
0   & 		0.1052
\end{array}
\right],
\end{align}
$C=[1 \,\,\,\,2]$ and $\Delta=0.1$s. The Laplacian matrix of the undirected and connected communication graph follows that in \cite{feng2020ifac}.
We select a feedback gain $K= \text{diag}(4.2, 4.2)$.

We consider a 
sustained DoS attack with variable period and duty cycle, generated randomly.
Over a simulation horizon of $10$s, the DoS 
signal yields $|\Xi(0,10)|=1.9$s and $n(0,10)=15$. 
This corresponds to values (averaged over $10$s) 
of $\tau_D\approx 0.6667$ and $1/T \approx 0.19$,
and hence $\Delta/\tau_D \approx 0.15$ and
$
\Delta/\tau_D + 1/T \approx 0.34
$.

Under this setting, one can obtain $\rho(J)=0.8146$, $\rho(A)=1.2731$, $C_J=1.1070$ and $C_A=1.0607$. Then we select $\gamma_1=0.85>\rho(J)$ and $\gamma_2=1.4 > \rho(A)$ according to Lemma \ref{Lemma3}. For the observer gain, we select $F=[0.2757\,\,
0.2134]^T$ with $\rho(A-FC)=0.81 < \rho(J)$. One can verify that $\gamma_0=0.9583$ (in \eqref{47}). One can also see that (\ref{bound Lemma 3}) holds in view of $\Delta/\tau_D \approx 0.15$ and $-\ln \gamma_{0} / \ln C_A C_J =0.3257$. By Theorem \ref{Theorem 1}, we obtain the right-hand side of (\ref{Theorem quantizer}) being 301920.

Simulation plots are presented in Figure \ref{simulation1}. We point out that the DoS frequency constraint (\ref{bound Lemma 3}) is satisfied in the simulation example, but the level of DoS attacks characterized by $
\Delta/\tau_D + 1/T \approx 0.34
$ is stronger than the theoretical sufficient bound computed by (\ref{DoS bound}), which is $0.3257$. Since our result regarding tolerable DoS attacks is a sufficient condition, one can see from the first and second plots in Figure \ref{simulation1} that state consensus is still achieved. When one increases $
\Delta/\tau_D + 1/T$ to about $0.4$, then the states $\delta_i^1(k)$ and $\delta_i^2(k)$ diverge.  
By the third plot in Figure \ref{simulation1}, one can see that $\theta(k)$ is a decreasing and increasing sequence during DoS-free and DoS time, respectively. Thanks to the dynamical $\theta(k)$, the quantizer is not overflowed, which can be seen from the last plot in Figure \ref{simulation1}. Specifically, the quantization range provided by the theoretical value $[-301920, 301920]$ is much larger than the utilized quantization range $[-5, 5]$ in simulation. One could see that our sufficient condition for quantizer unsaturation is quite conservative. 
One of the reasons is that we have frequently used ``$\le$" and $``\max\{\cdot\}"$ in (\ref{upper bound}) for instance and $\|\cdot\|_\infty$ for matrices and vectors in (\ref{61}) for instance. Moreover, a small $\gamma_1$ can also lead to a large data rate as discussed in \cite{feng2020arxiv}. Without DoS, such a conservativeness also exists in consensus under data rate limitation in \cite{you2011network, li2010distributed} for instance.

\emph{Scalar multi-agent systems:}
We consider the multi-agent system in the numerical example in \cite{you2011network}, in which $A=1.1$, $B=1$ and $N=4$. The Laplacian matrix of the undirected and connected communication graph follows that in the previous example.
We select $K=0.44$ and $\Delta = 0.1$s. One has $\rho(J)=0.66$. According to Proposition \ref{Theorem 2}, we choose $\gamma_1=0.67$ and $\gamma_2=1.0962$, and quantizer parameter $(2R + 1)\sigma$ should be no smaller than $183890$. Besides, the sufficient DoS condition for consensus is 
\begin{align}\label{72}
\!\!\!1/T + \Delta / \tau_D  &< \frac{-\ln \gamma_1}{\ln \gamma_2 \!- \! \ln \gamma_1} =\!   \frac{- \ln (\rho(J)/ \gamma_1)  }{\ln (A/\gamma_2)  - \ln (\rho(J)/ \gamma_1) } \nonumber\\ &\! =\!
\frac{-\ln \rho(J)}{\ln A - \ln \rho(J)} 
=  0.8134.
\end{align}

\begin{figure}[t]
	\begin{center}
		\vspace{-4mm}
		\includegraphics[width=0.49\textwidth]{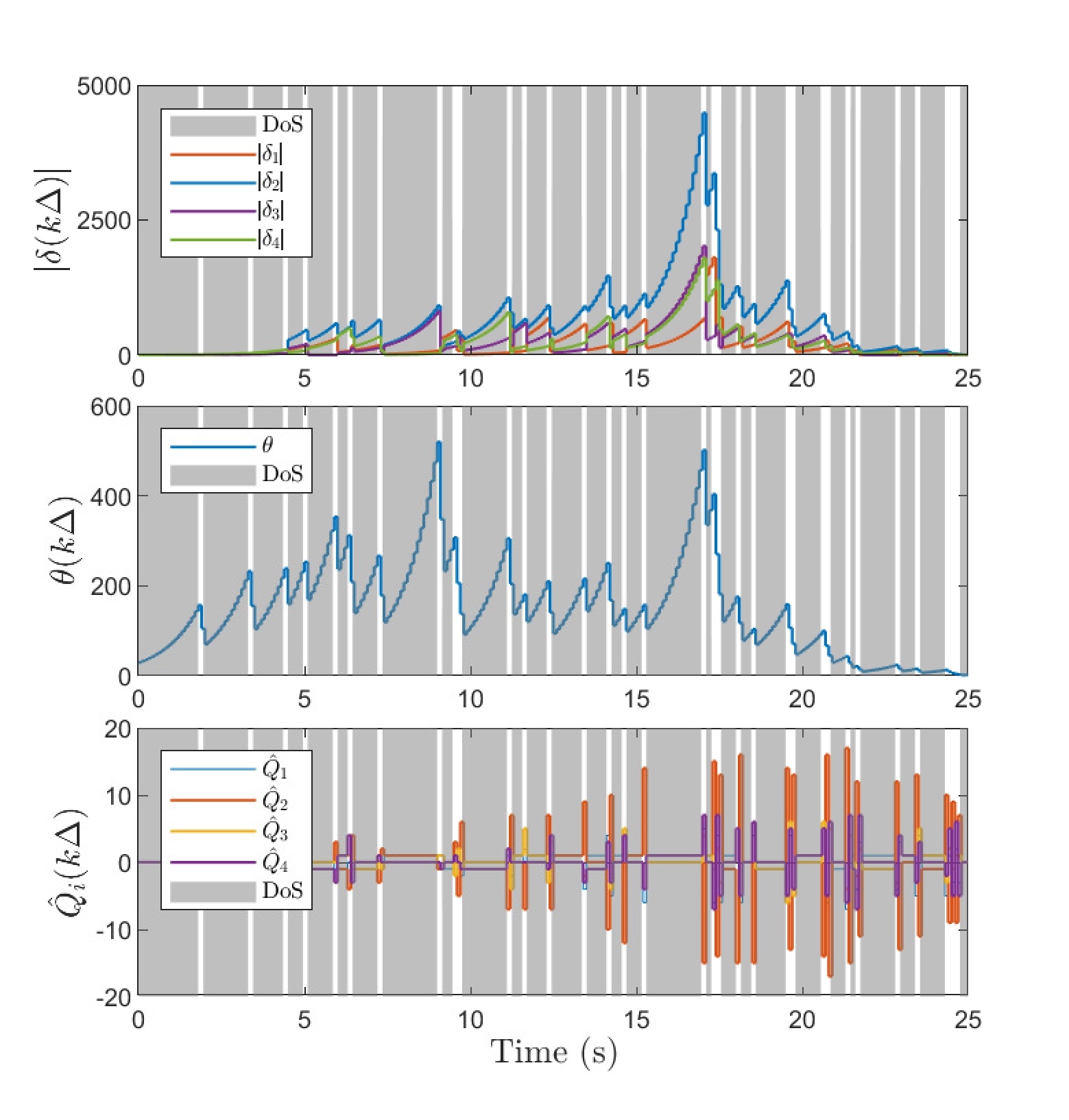}  \\
					\vspace{-5mm}
		\linespread{1}\caption{Time responses of $|\delta_i(k)|$ (top), $\theta(k)$ (middle) and $\hat Q _i (k)$ (bottom). } \label{simulation}
	\end{center}
\vspace{-5mm}
\end{figure}

Similar to the previous simulation example, the randomly generated DoS
over a simulation horizon of $25$s (gray stripes in Figure \ref{simulation}) yields $|\Xi(0,25)|=20.5$s and $n(0,25)=28$. 
This corresponds to values (averaged over $25$s) 
of $\tau_D\approx 0.8929$ and $T\approx1.2195$,
and the DoS attacks in this example yield
$
\Delta/\tau_D + 1/T \approx 0.9320
$. 


The simulation results are presented in Figure \ref{simulation}. The convergence of $|\delta_i(k)|$ ($k=1, 2, 3, 4$) is presented in the first plot of Figure \ref{simulation}, in which one can see $|\delta_i(k)|$ increases during DoS intervals (gray areas) and decreases when DoS is not present (white areas). Note that  $|\delta_i(k)|\to 0$ implies the state consensus. The zooming-in and zooming-out mechanism can be observed by the second plot in Figure \ref{simulation}, in which $\theta(k)$ increases and decreases during DoS present and absent intervals, respectively. The effectiveness of zooming-in and out mechanism is shown by the third plot of Figure \ref{simulation}. Though the state increases during DoS present intervals, with the zooming out of $\theta(k)$ for mitigating the influence of DoS, one can see that the actual value of $\hat Q_i(k)$ does not diverge under DoS. 
\color{black}  
Importantly, compared with the $\gamma_2=4.0333$ and the DoS bound 0.2037 in \cite{feng2020ifac}, one can see that the zooming-out factor $ 1.0962<A$ in this paper and the bound for tolerable DoS 0.8134 in (\ref{72}) are indeed much improved. 

Conservativeness also exists in the case of scalar multi-agent systems, which can be seen by the gaps of DoS level between 0.8134 (theoretical sufficient bound in (\ref{72})) and 0.9320 (actual DoS in the simulation), and the gaps between the theoretical quantization range $[-183890, 183890]$ and the utilized range $[-19, 19]$ shown in the third plot of Figure \ref{simulation}.

\section{Conclusions and future research}

We have presented the results for quantized consensus of output feedback multi-agent systems. The design of dynamic quantized controller including the observer and the zooming-in and zooming-out parameters has been presented. The calculation of zooming-out factor is tight, whose lower bound is the spectral radius of the agent's dynamic matrix. Moreover, the approach of this paper shows the explicit relations between the zooming-out factor and the agent dynamic matrix.
We have also provided the bounds of quantizer range and tolerable DoS attacks. 
It has been shown that the quantizer is free of overflow during DoS intervals and the state consensus can be achieved. At last, as a special case of scalar multi-agent systems, we have shown 
that it is possible to further tighten the zooming-out factor and make it smaller than the agent's system parameter without causing quantizer overflow. The resilience is also improved by such a zooming-out factor, i.e., recover to that of unquantized consensus under DoS.

Our work can be extended to various directions. Following \cite{meng2016coordination}, one can consider the control structure in which each agent directly exchanges output information with the neighbors by implementing a modified observer in the decoder. 
It is possible to extend our results to a directed graph of the communication topology, in which the analysis after Section III-C will need to be adapted \cite{8757723}. Last but not least, it is meaningful to consider the case in which the decoded values of a state are different due to communication noise or parameter uncertainties for instance.

\appendix

\emph{Proof for (\ref{optimal}).}
In case $A\in \mathbb R$ and $C \in \mathbb R$, it is easy to obtain $x_i(k)$ directly from $y_i(k) = Cx_i(k)$ and an observer is not necessary. Hence, we assume that $y_i(k)=x_i(k)$. In case of infinite data rate, one can obtain that 
$
\delta(k) = 
\left\{\!\!\!\!
\begin{array}{ll}
G\delta(k-1) & \text{if}\, k-1 \notin H_q \\
A_N\delta(k-1) & \text{if}\,k-1 \in H_q
\end{array}
\right.\!\!\! \!\!  .
$
With the unitary matrix $U$ in (\ref{Phi}) and by $\tilde \delta(k) = U ^ T \delta (k)$,  one can obtain that 
$
\tilde \delta(k) \!\!=\!\!  
\left\{\!\!\!
\begin{array}{ll}
D   \tilde \delta(k-1) & \text{if}\, k-1 \notin H_q \\
A_N \tilde \delta(k-1) & \text{if}\,k-1 \in H_q
\end{array}
\right.\!\!\!\!\!,
$
in which $D:= \text{diag}(A, A-\lambda_2BK, ..., A- \lambda_N BK)$.
Partition the vector $\tilde \delta(k)= [\tilde \delta_1 (k) \,\, \tilde \delta_2 ^ T (k) ] ^ T $, in which $\tilde \delta_1 (k) \in  \mathbb R $ is the first component in the vector and $\tilde \delta_2 (k) \in  \mathbb R ^ {N-1}$ is composed by the rest. Then we obtain the dynamics of $\tilde \delta_2 (k)$ as
$
\tilde \delta _ 2 (k)\!\! =\!\!  
\left\{\!\!\!
\begin{array}{ll}
J   \tilde \delta _ 2 (k-1) & \text{if}\, k-1 \notin H_q \\
A_{N-1} \tilde \delta _ 2 (k-1) & \text{if}\,k-1 \in H_q
\end{array}
\right.\!\!\!\!
$
and therefore
$
\|\tilde \delta _ 2 (k)\| \!\!\le\!\! 
\left\{\!\!\!\!
\begin{array}{ll}
\rho(J) \|\tilde \delta _ 2 (k-1)\| & \text{if}\,\, k-1 \notin H_q \\
A \|\tilde \delta _ 2(k-1)\| & \text{if}\,\, k-1 \in H_q 
\end{array}
\right.\!\!\!.
$
By the iteration of $\|\tilde \delta _ 2 (k)\|$, one can obtain $\|\tilde \delta_2 (k)\| \le A^ {T_U(1,k)} \rho(J) ^{T_S (1,k)} \|\tilde \delta_2(1)\| $. 
Note that $\tilde \delta_1 (k) = 0$ for all $k$. Then one can obtain $\|\tilde \delta_2 (k) \| = \|\tilde \delta(k)\|$ and $\|\delta(k) \| = \|   U \tilde \delta(k)\|  = \|\tilde \delta(k)\|$.
By substituting the bound in Lemma \ref{Lemma T} into $T_S(1, k)$ and $T_U(1,k )=k-T_S(1, k)$, one can obtain that 
$
\|\delta(k)\| \le  C_U  \gamma_U  ^k   \|\delta (1)\| 
$
where $C_U  = \left( A / \rho(J)  \right)^{(\kappa+\eta\Delta)/\Delta} $. 
If the level of DoS attacks satisfy (\ref{optimal}), then
$
\gamma _ U : = \rho(J) ^{1-\frac{1}{T}-\frac{\Delta}{\tau_D}  } A ^ { \frac{1}{T}+ \frac{\Delta}{\tau_D}   } < 1.
$
As $k \to \infty$, one has $\|\delta(k)  \| \to 0 $, which implies the consensus of $x_i(k)$.  
\qedp

\bibliographystyle{IEEEtran}

\bibliography{multiquantization}

\end{document}